\documentclass[twocolumn,aps,prl,superscriptaddress]{revtex4-2}

\usepackage[Symbol]{upgreek}
\usepackage{stix}
\usepackage{graphicx}
\usepackage{bm}
\usepackage{amsmath}
\usepackage{color}

\newcommand{\dd}{\mathrm{d}}
\newcommand{\ii}{\mathrm{i}}

\newcommand{\tm}{\tilde{m}}

\newcommand{\bb}{\bm{b}}

\begin{document}

\title{Stacking Domain Wall Magnons in Twisted van der Waals Magnets}

\author{Chong \surname{Wang}}
\affiliation{Department of Physics, Carnegie Mellon University, Pittsburgh, Pennsylvania 15213, USA}

\author{Yuan \surname{Gao}}
\affiliation{Department of Physics, Carnegie Mellon University, Pittsburgh, Pennsylvania 15213, USA}
\affiliation{International Center for Quantum Design of Functional Materials (ICQD), Hefei National Laboratory for Physical Sciences at the
Microscale, University of Science and Technology of China, Hefei, Anhui 230026, China}

\author{Hongyan \surname{Lv}}
\affiliation{Department of Physics, Carnegie Mellon University, Pittsburgh, Pennsylvania 15213, USA}
\affiliation{Key Laboratory of Materials Physics, Institute of Solid State Physics, Chinese Academy of Sciences, 
Hefei, 230031, China}

\author{Xiaodong \surname{Xu}}
\affiliation{Department of Physics, University of Washington, Seattle, Washington 98195, USA}
\affiliation{Department of Materials Science and Engineering, University of Washington, Seattle, Washington 98195, USA}

\author{Di \surname{Xiao}}
\affiliation{Department of Physics, Carnegie Mellon University, Pittsburgh, Pennsylvania 15213, USA}

\begin{abstract}
Using bilayer CrI$_3$ as an example, we demonstrate that stacking domain walls in van der Waals magnets can host one dimensional (1D) magnon channels, which have lower energies than bulk magnons. Interestingly, some magnon channels are hidden in magnetically homogeneous background and can only be inferred with the knowledge of stacking domain walls. Compared to 1D magnons confined in \emph{magnetic} domain walls, 1D magnons in \emph{stacking} domain walls are more stable against external perturbations. We show that the relaxed moir\'e superlattices of small-angle twisted bilayer CrI$_3$ is a natural realization of stacking domain walls and host interconnected moir\'e magnon network. Our work reveals the importance of stacking domain walls in understanding magnetic properties of van der Waals magnets, and extends the scope of stacking engineering to magnetic dynamics.
\end{abstract}

\maketitle

The physical properties of two-dimensional (2D) van der Waals (vdW) materials depend sensitively on the stacking arrangement between adjacent layers.  Consequently, the modulation of stacking  can strongly modify the local electronic properties. For example, in bilayer graphene, the stacking domain walls between the AB and BA stackings~\cite{alden_strain_2013,butz2014} induce a variation in the electronic Hamiltonian, which can give rise to one dimensional (1D) topologically protected electronic states~\cite{zhang2013,vaezi2013,ju_topological_2015,yin_direct_2016}. Similarly, in transition metal dichalcogenides (TMD), electrons can be confined in stacking domain walls, which can be controlled experimentally via strain engineering~\cite{edelberg_tunable_2020}.

Recently it has been realized that the stacking dependence also extends to magnetic properties. For example, in bilayer CrI$_3$, the interlayer exchange coupling changes sign as the stacking is varied~\cite{sivadas_stacking-dependent_2018,wang2018,jiang_stacking_2019,soriano2019,jang2019,chen2019direct}.  Therefore a stacking domain wall also induces a modulation in the spin Hamiltonian.  It is thus natural to expect stacking domain walls to host 1D spin wave (magnon) channels. Previously, confined 1D magnons have been proposed to exist in \emph{magnetic} domain walls~\cite{winter_bloch_1961,thiele_excitation_1976,ferrer_one-dimensional_2003,Sanchez_narrow_2015,lan_spin-wave_2015}.  However, magnetic domain walls are generally fragile with respect to external perturbations and may even have its own dynamics. On the other hand, \emph{stacking} domain walls, whose energy scale is at least one order of magnitude larger than magnetic domain walls, provide a more stable platform to host 1D magnons. 

\begin{figure}
\centering
\includegraphics[width=1.0\columnwidth]{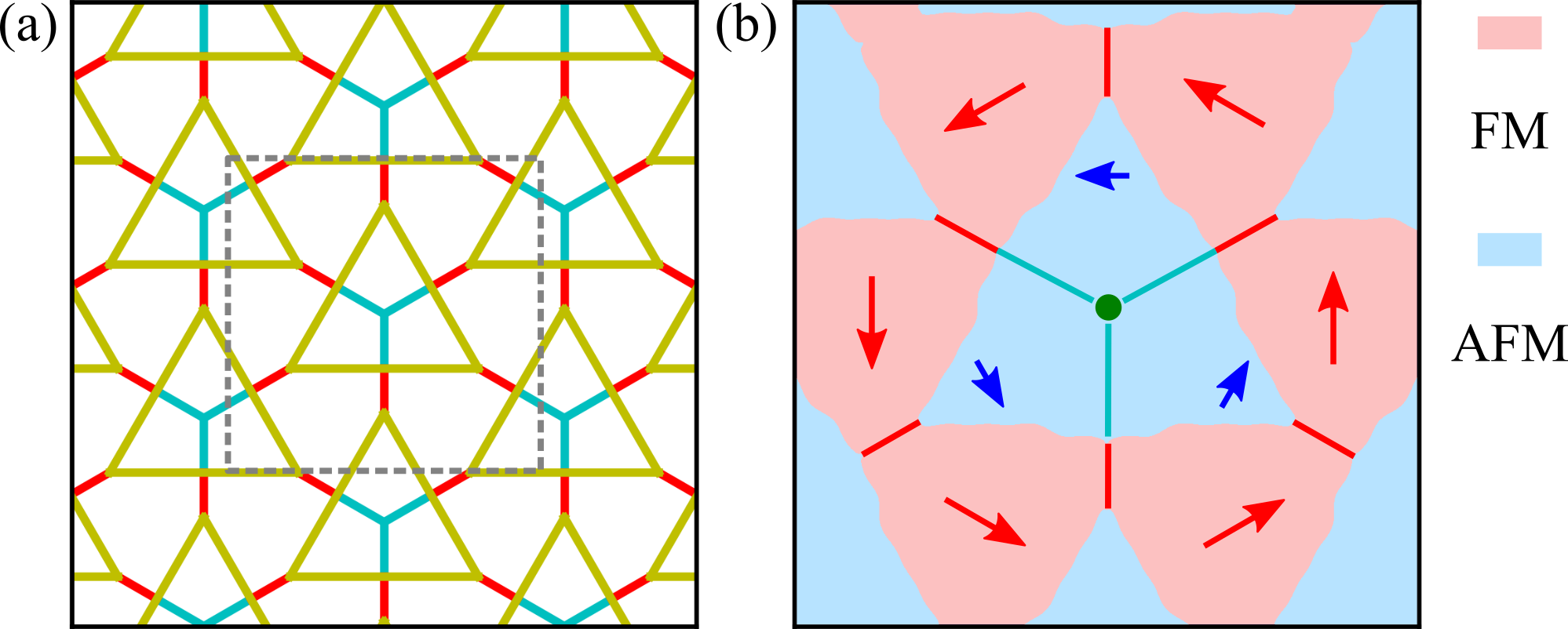}
\caption{{(a) Sketch of the magnon network in twisted bilayer CrI$_3$ with twist angle $0.1^{\circ}$.} Panel (b) zooms into the region marked by the grey rectangle in panel (a) and shows the stacking and magnetic domain patterns. Red (blue) arrows represent stacking vectors for rhombohedral (monoclinic) stackings. Red (cyan) lines represent the RR (MM) stacking domain walls.\label{fig:moire}}
\end{figure}

In this work, using bilayer CrI$_3$ as an example, we study 1D magnons in stacking domain walls in vdW magnets. We show that, quite generally, all stacking domain walls of bilayer CrI$_3$ can host 1D magnons,  which have lower energies than bulk magnons. The existence of these 1D magnons can be adiabatically traced back to the Goldstone modes of the spin Hamiltonian. Interestingly, we find that some magnon channels are hidden in magnetically homogeneous background and can only be inferred with the knowledge of stacking domain walls. These domain walls are naturally realized in moir\'e superlattices in twisted bilayer magnets with small twist angles. Moir\'e magnons have been recently studied in Refs.~[\onlinecite{hejazi2020moir,li_moire_2020,ghader_magnon_2020}]. However, these works have ignored lattice relaxation, and the information of stacking domain walls are not utilized in the construction of the spin Hamiltonian. With a full account of lattice relaxation, we calculate the stacking and magnetic moir\'e pattern in small angle twisted bilayer CrI$_3$ (Fig.~\ref{fig:moire}).  In this system, the stacking domain walls and corresponding 1D magnons are interconnected, forming a magnon network, which will dominate low-energy spin and thermal transport. Our work reveals the importance of stacking domain walls in understanding magnetic properties of vdW magnets, and extends the scope of stacking engineering to magnetic dynamics.

\emph{Stacking domain wall}.---CrI$_3$ is a layered magnetic material whose magnetic order can survive down to the monolayer limit~\cite{huang_layer-dependent_2017}. The Cr atoms in a monolayer CrI$_3$ forms a hexagonal lattice with lattice constant 6.9 \AA. Monolayer CrI$_3$ is a ferromagnet with an out-of-plane easy axis. Bilayer CrI$_3$ has two stable stackings, i.e., rhombohedral and monoclinic stackings. Both stackings have roughly the same energy, but rhombohedral stacking strongly favors interlayer ferromagnetic (FM) configuration while monoclinic stacking weakly favors interlayer antiferromagnetic (AFM) configuration~\cite{sivadas_stacking-dependent_2018,wang2018,jiang_stacking_2019,soriano2019,jang2019,chen2019direct}.

\begin{figure}
\centering
\includegraphics[width=1\columnwidth]{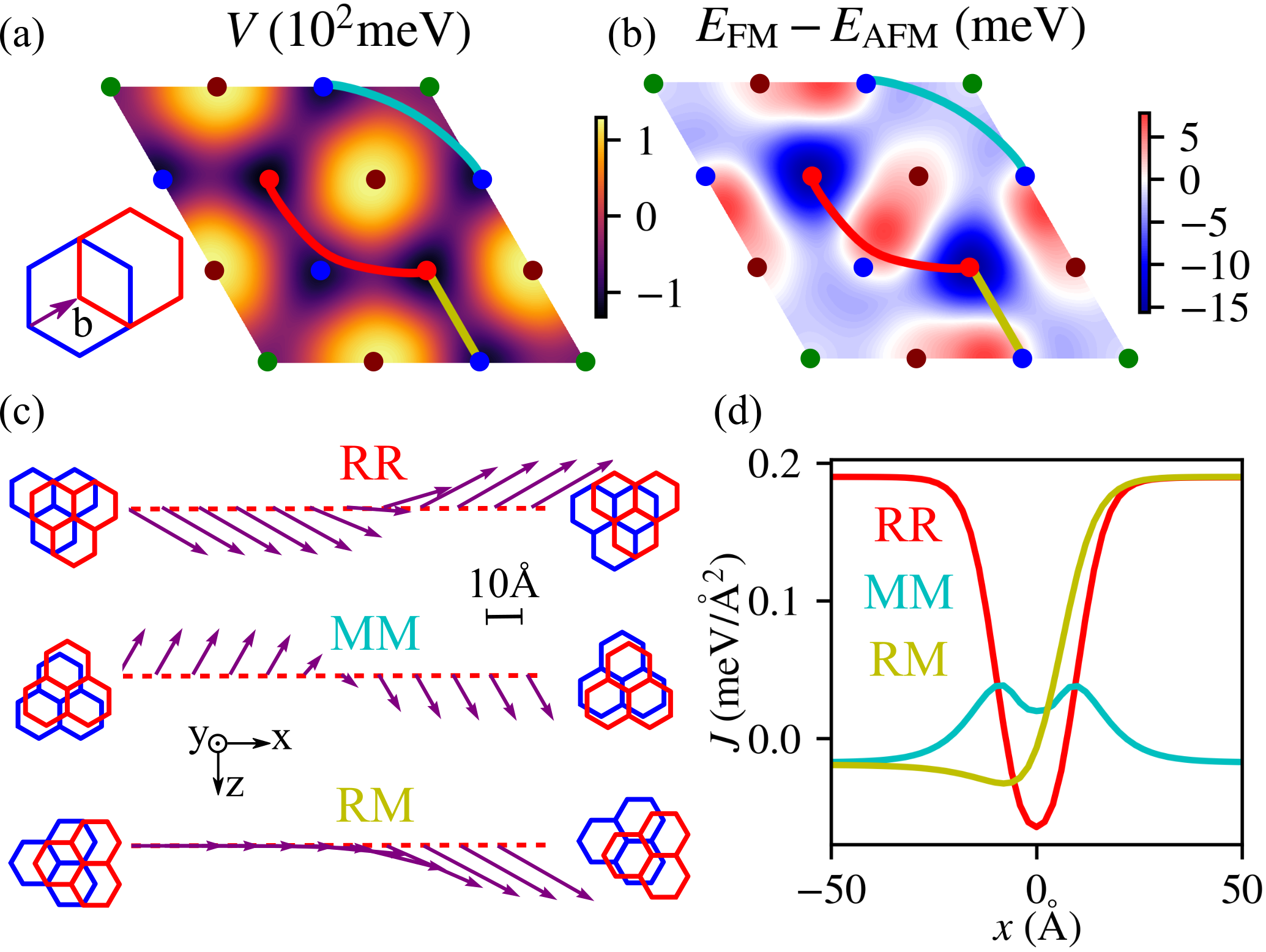}
\caption{(a) Interlayer potential energy per unit cell as a function of stackings $\bb$ for bilayer CrI$_3$ calculated with density functional theory~\cite{supplemental}. Important stackings are rhombohedral (red dot), monoclinic (blue dot), AA (green dot) and AC (brown dot) stackings. Red, cyan and yellow lines represent the paths of the stacking vector $\bb$ for RR, MM and RM domain walls, respectively. Inset: definition of the stacking vector $\bm{b}$; red (blue) lattice denotes the top (bottom) layer. (b) The energy difference per unit cell between interlayer FM and interlayer AFM CrI$_3$ as a function of stackings. (c) The top view of the three types of stacking domain walls, with the stacking vectors denoted by {purple} arrows. The stacking of the left (right) domain is sketched on the left (right) side. (d) $J(x)$ across different types of stacking domain walls.\label{fig:domainwall}}
\end{figure}

The stacking configuration is described by the relative in-plane displacement $\bb$ between the top and bottom layer [Fig. \ref{fig:domainwall}(a)]. $\bb$ is only defined modulo lattice translations. All possible values of $\bb$ constitute the stacking space, which coincides with the unit cell of {monolayer} CrI$_3$. {We first consider a 1D domain wall separating two semi-infinite planes with different stackings,} which are described by a continuous variation of the stacking vector $\bb$. In our study of each domain wall, we choose our coordinates such that the domain wall always lies in the $z$ direction at $x\sim 0$. {The $y$ axis is the easy axis, pointing to the out-of-plane direction.} $x \ll 0$ is the left stacking domain, described by $\bb(-\infty)=\bb_\mathrm{left}$. Similarly, $\bb(+\infty)=\bb_\mathrm{right}$ is the stacking vector of the right stacking domain. Around $x=0$, $\bb(x)$ changes rapidly from $\bb_\mathrm{left}$ to $\bb_\mathrm{right}$, passing through a series of unstable stackings. In CrI$_3$, domain walls can appear between two rhombohedral stackings (denoted as RR domain wall), between rhombohedral and monoclinic stackings (RM domain wall) and between two monoclinic stackings (MM domain wall).

To quantitatively characterize $\bb(x)$ for each domain wall, we employ continuous elastic theory. Since the energy scale of different stackings is larger than the energy scale of magnetism by at least one order of magnitude, we ignore the influence of magnetism at this stage. The stacking energy of the domain wall is a summation of the interlayer potential energy and the elastic energy:
\begin{equation}
    E_{\mathrm{str}} = \frac{1}{|P|}\int_{-\infty}^\infty \Bigl[V(\bb) + \frac{B + G}{4} (\partial_x b^x)^2 + \frac{G}{4} (\partial_x b^z)^2 \Bigr] \dd x, \label{energy_str}
\end{equation}
where $V(\bb)$ is interlayer potential energy per unit cell [Fig.~\ref{fig:domainwall}(a)], $|P|$ is the area of the unit cell, $B=54307$ meV per unit cell is the bulk modulus of monolayer CrI$_3$ and $G=39248$ meV per unit cell is the shear modulus~\cite{supplemental}. In Eq.~\eqref{energy_str}, we have assumed that the absolute displacements of the top and the bottom layer are $\pm\bb(x)/2$.

The elastic energy $E_{\mathrm{str}}$ can be minimized by solving the Euler-Lagrangian equation $\delta E_\mathrm{str}/\delta \bb = \bm{0}$. For the RM domain wall, $V$ is roughly reflection symmetric with respect to the line connecting rhombohedral and monoclinic stackings [yellow line in Fig.~\ref{fig:domainwall}(a)], and we can safely assume $\bb(x)$ simply takes this straight path [$b^x(x)=0$]. Furthermore, $V$ can be approximated as a cosine function on the path of the stacking vector: the two minima of the cosine function at $\pm \uppi$ correspond to rhombohedral and monoclinic stackings; the maximum of the cosine function is in the middle between the two stable stackings. With this assumption, {the Euler-Lagrangian equation reduces to the sine-Gordon equation and admits a soliton solution}:
\begin{equation}
    b^z = 2(b_\mathrm{right}^z-b_\mathrm{left}^z) \arctan [ \exp (x/w) ]/\uppi + b_\mathrm{left}^z.\label{soliton}
\end{equation}
The characteristic width $w=|\bb_\mathrm{left}-\bb_\mathrm{right}|\sqrt{G/V_0}/\uppi$ is roughly 8.8 \AA. Here, $V_0$ is the barrier of $V$ along the path of $\bb(x)$. Notice that the range of $x$ for which $\arctan [ \exp (x/w) ]$ varies significantly is roughly $6w$.

In contrast, for the RR and MM domain walls, $\bb(x)$ has to bypass a high energy barrier [brown dot in Fig.~\ref{fig:domainwall}(a)], and analytic solutions cannot be obtained. We instead numerically solve the Euler-Lagrangian equation and present $\bb(x)$ as the red line and cyan line in Fig.~\ref{fig:domainwall}(a) for the two types of domain walls. Nevertheless, it is possible to fit $b^z$ to Eq.~\eqref{soliton} and the characteristic widths for the RR and MM domain walls are 9.5 and 7.5~\AA, respectively. All three types of stacking domain walls are plotted in Fig.~\ref{fig:domainwall}(c). Note that we are studying shear domain walls and therefore $\bb_\mathrm{left}-\bb_\mathrm{right}$ is always along the $z$ direction.

\begin{figure}
\centering
\includegraphics[width=1\columnwidth]{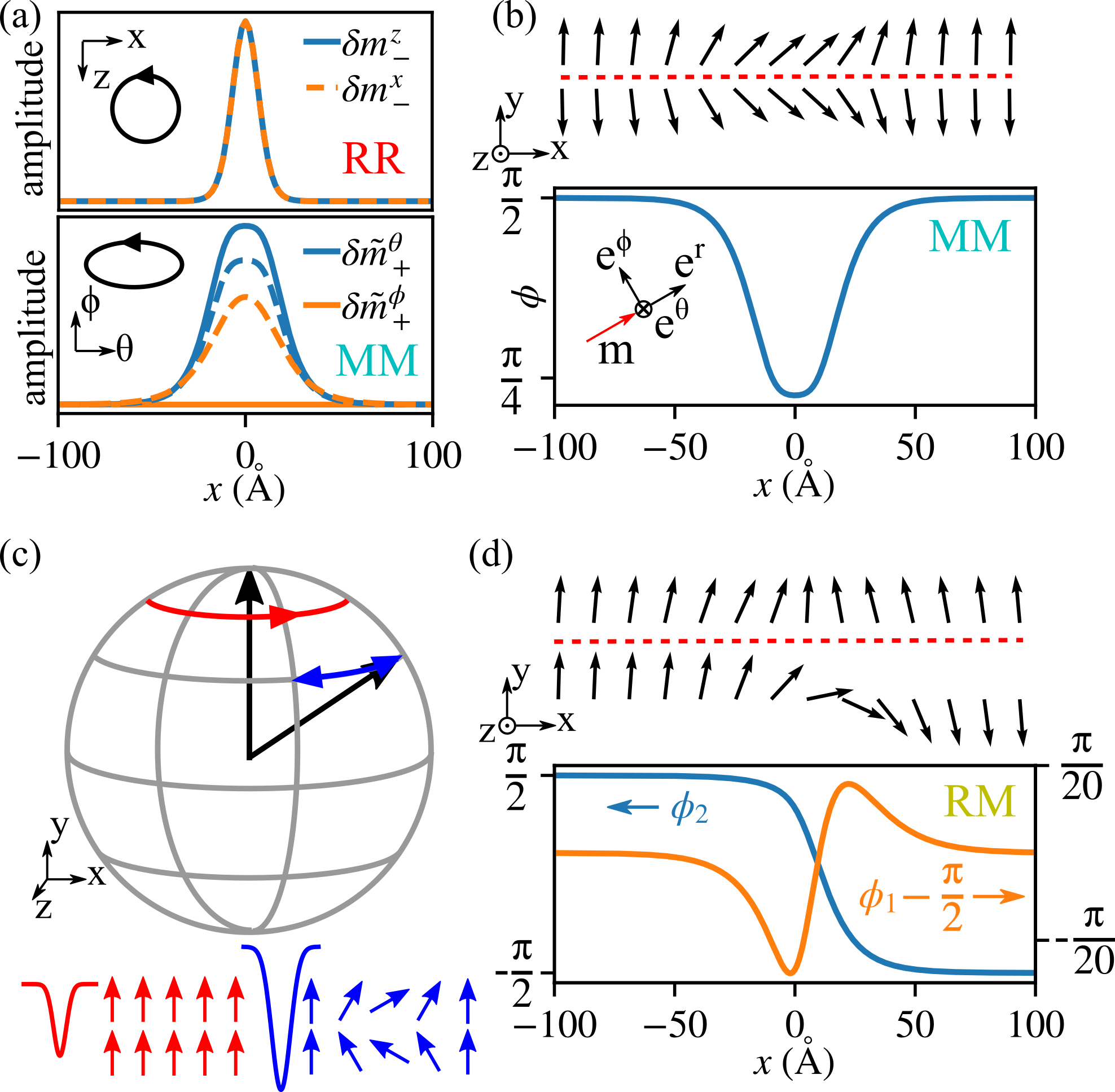}
\caption{(a) Bounded magnons in the RR (top) and MM (bottom) domain walls. For the bottom panel, solid (dashed) lines correspond to $k_z=0.0$ ($0.05$) 1/\AA. Insets: motions of the magnons. (b) Magnetic ground state of MM domain wall. Top panel: magnetization; bottom panel: azimuth angles of the top layer magnetization. Bottom panel inset: spatial varying coordinates. $\hat{\bm{e}}^{r}$ is along the direction of $\bm{m}$; ($\hat{\bm{e}}^{\theta}$, $\hat{\bm{e}}^{\phi}$) are defined as the unit vectors in the direction of increasing $\theta$ and $\phi$. (c) Schematic plots of confined magnons (top) and magnetic ground states (bottom) for weak (red) and strong (blue) trapping potentials $J(x)$. The top panel is a Bloch sphere representing the direction of magnetization. For weak trapping potential, the magnon is a circular motion around the north pole ($y=1$); for strong trapping potential, the confined magnon is a Goldstone mode oscillating on the latitude line (constant $y$). (d) Magnetic ground state of the RM domain wall. Azimuth angles of the magnetization of the top layer ($\phi_1$) and bottom layer ($\phi_2$) are plotted in the bottom panel.\label{fig:magnon}}
\end{figure}

\emph{1D magnon channel}.---The variation of the stacking vector $\bb$ induces a variation in interlayer exchange coupling. In Fig.~\ref{fig:domainwall}(b), we plot interlayer exchange couplings for different stackings together with paths of $\bb(x)$ for the three types of stacking domain walls. To study the magnetic properties of the domain walls, we {focus on the long-wavelength continuum limit} and adopt the following micromagnetics energy
\begin{equation}
    E_{\mathrm{mag}} = \int \Bigl[\sum_{\alpha, \beta, l} \frac{A}{2} (\partial_{\beta} m_l^{\alpha})^2 - \sum_l \frac{K}{2}  (m_l^y)^2 - \sum_{\alpha} J m_1^{\alpha} m_2^{\alpha}\Bigr] {\dd^2 \bm{r}},\label{mag_energy}
\end{equation}
where $\alpha$, $\beta$ are Cartesian indices, $l$ is layer index and $\bm{m}$ is the unit vector pointing in the direction of magnetization. The first term and the second term in $E_{\mathrm{mag}}$ is intralayer FM coupling and magnetic anisotropy, where $A\approx 5.3$ meV and $K\approx 0.032$ meV/\AA$^2$ for monolayer CrI$_3$. The last term describes the interlayer exchange coupling, where $J$ depends on $x$. {The value of $J(x) \equiv J(\bm b(x))$ is extracted from Fig.~\ref{fig:domainwall}(b) and is plotted in Fig.~\ref{fig:domainwall}(d) for different domain walls.}

We start with the magnetic properties of the RR domain wall. Rhombohedral stacking strongly favors interlayer FM configuration. In the domain wall, the stacking vector passes through an region favoring interlayer AFM configuration [Fig.~\ref{fig:domainwall}(b, d)]. However, this interlayer AFM tendency is punished by both intralayer FM coupling ($\propto A$) and the magnetic anisotropy ($\propto K$). To quantitatively characterize the competition, we carry out simulations of the Landau-Lifshitz-Gilbert equation with the damping term. Our simulation shows that the weak interlayer AFM coupling within the domain wall does not change the direction of the magnetization and the ground state is simply described by a uniform, out-of-plane $\bm m$.

The variation of interlayer exchange coupling, although not manifested in the magnetic ground state, will enter the equation of motion of magnetization dynamics, $\partial_t\bm{m}_l = -\gamma\bm{m}_l\times \bm{H}_l$, where the effective magnetic field is $\bm{H}_l = A \nabla^2 \bm{m}_l + K m^y \hat{\bm{y}} + J \bm{m}_{\bar{l}}$ ($\bar{1}=2$, $\bar{2}=1$ and $\gamma$ is the gyromagnetic ratio). Since $\bm{m}$ is a unit vector, the first order variation of magnetization is written as $\delta\bm{m}_l=\delta m_l^x \hat{\bm{x}}+\delta m_l^z \hat{\bm{z}}$. The ground state respects mirror symmetry in the $y$ direction ($M_y$), so it is natural to decouple the mirror eigenspaces by defining $\delta \bm{m}_\pm = (\delta \bm{m}_1 \pm \delta \bm{m}_2)/2$. With the notation $\delta m_\pm^+=\delta m_\pm^x+\ii \delta m_\pm^z$, the dynamical equation can be transformed into two decoupled Schr\"odinger type equations:
\begin{equation}
\begin{aligned}
\ii \gamma^{-1} \partial_t \delta m_+^+ &= [- A \partial_x^2 + K + A k_z^2] \delta m_+^+, \\
\ii \gamma^{-1} \partial_t \delta m_-^+& = [- A \partial_x^2 + K + 2J(x) + A k_z^2] \delta m_-^+,
\end{aligned}\label{RRMagnon}
\end{equation}
where we have assumed $\boldsymbol{m}_l$ behaves like a plane wave in the $z$ direction with wave vector $k_z$. We see that $\delta \bm{m}_+$ is blind to the interlayer coupling. For $\delta \bm{m}_-$, since the variation of $J(x)$ serves as a trapping potential [Fig.~\ref{fig:domainwall}(d)], despite the magnetization of the ground state is uniform across the RR domain wall, 1D magnon solutions generally exist and are confined in the domain wall. In Fig.~\ref{fig:magnon}(a) we present the profile of the 1D magnon with the lowest energy. It is a circular motion isotropic in the $x-z$ plane and the magnon profile is independent of $k_z$. {The frequency of this 1D magnon mode at $k_z=0$ is calculated to be $0.3 \gamma K$. Here $\gamma K$ is the bulk magnon frequency gap and is experimentally measured~\cite{cenker_direct_2020} as $\sim 0.3$ meV, from which we deduce the energy of the 1D magnon is $\sim 0.09$ meV.} Therefore, the low energy magnon transport should be dominated by this 1D magnon channel.

To gain more insights into the existence of this 1D magnon, we assume the width and depth of the trapping potential $J(x)$ can be artificially tuned. For a weak trapping potential, the magnetic ground state is a uniform interlayer FM configuration across the RR domain wall [Fig.~\ref{fig:magnon}(c)], which is the case for CrI$_3$. The degree of freedom $\delta \bm{m}_-$ describes the deviation from interlayer FM configuration and the frequency of the corresponding magnon is determined by the energy cost of such deviation. Since in the domain wall the interlayer exchange coupling $J(x)$ becomes smaller, such energy cost is lower in the domain wall. Therefore, a confined mode should exist in the domain wall, which is manifested as a bound state due to $J(x)$ in Eq.~\eqref{RRMagnon}. For increasingly stronger trapping potential, there will be a critical point where the frequency of this 1D magnon at $k_z = 0$ becomes zero. Beyond this critical point, the magnetic ground state deviates from the interlayer FM configuration in the domain wall [Fig.~\ref{fig:magnon}(c)]. However, the 1D magnon mode does not disappear. The magnetic energy described by Eq.~\eqref{mag_energy} is actually invariant under a global (independent of $x$) rotation about the $y$ axis. This continuous symmetry immediately gives rise to a Goldstone magnon mode, which costs no energy at $k_z=0$ and is still localized in the domain wall [Fig. \ref{fig:magnon}(c)]. It is worth noting that arbitrarily weak trapping potentials can confine 1D magnons. However, if $J(x)$ is always positive, the frequency of the 1D magnon will be larger than the bulk magnon gap $\gamma K$.

Similar to the RR domain wall, the MM stacking domain wall appears between two magnetically identical (AFM) domains. However, since the AFM interlayer coupling in CrI$_3$ is much weaker than the FM interlayer coupling, the magnetization now tilts away from the $y$-axis in the MM stacking domain wall. We parameterize the magnetization of the top layer $\bm{m}_1(x)$ in spherical coordinates $\theta$ (polar angle) and $\phi$ (azimuth angle) and present the magnetic ground state in Fig.~\ref{fig:magnon}(b). The Goldstone mode argument discussed above immediately gives rise to a zero-energy 1D magnon mode trapped in the MM domain wall. Nevertheless, it is instrumental to write down the magnetic dynamical equations. We expand $\delta \bm{m}_l$ as $\delta \bm{m}_l = \delta m_l^\theta \hat{\bm{e}}_l^\theta + \delta m_l^\phi \hat{\bm{e}}_l^\phi$, where ($\hat{\bm{e}}_l^\theta$, $\hat{\bm{e}}_l^\phi$) is the local unit vectors associated with $\bm{m}_l$ [Fig.~\ref{fig:magnon}(b)]. Instead of $M_y$, the ground state now respects two fold rotation symmetry in the $x$ direction ($C_{2x}$). We decouple the eigenspaces of $C_{2x}$ by defining $\delta \tm^{\theta(\phi)}_\pm = (\delta m^{\theta(\phi)}_1 \pm \delta m^{\theta(\phi)}_2)/2$. The dynamical equation for $\delta \bm{\tm}_+$ is
\begin{equation}
\begin{aligned}
\gamma^{-1}\partial_t \tm_+^\theta &= (A \partial_x^2 - U^\phi - A k_z^2) \delta \tm_+^\phi \;, \\
\gamma^{-1}\partial_t \tm_+^\phi &= (- A \partial_x^2 + U^\theta + A k_z^2) \delta \tm_+^\theta \;,
\end{aligned}\label{MMmagnon}
\end{equation}
where $U^\phi = - K \cos(2\phi)$ and $U^\theta = -A (\partial_x \phi)^2 + K \sin^2 (\phi) - J[1 - \cos(2\phi)]$. Every term in $U^\phi$ and $U^\theta$ is a trapping potential, which is a combined effect of varying magnetization and varying $J(x)$. These trapping potentials serve as an alternative explanation for the 1D {zero-energy} Goldstone magnon mode in the MM domain wall. The profile of the lowest-energy 1D magnon of $\delta \bm{\tm}_+$ is shown in Fig.~\ref{fig:magnon}(a). For $k_z=0$, the magnon is perfectly polarized: $\delta \tm^\phi_+=0$, consistent with the Goldstone mode oscillating around the $y$ axis. This perfect polarization is reduced by finite $k_z$ [Fig.~\ref{fig:magnon}(a)]. Equation~\eqref{MMmagnon} shows that even when $\phi=\uppi/2$ across the domain wall [e.g., for weaker variation of $J(x)$] and the Goldstone mode argument fails, MM domain walls can still support 1D magnons due to the variation of $J(x)$ alone.

For $\delta \bm{\tm}_-$, whether the variation of $J$ and $\phi$ serves as a trapping potential depends on the specific parameters. Nevertheless, for bilayer CrI$_3$, $\delta \bm{\tm}_-$ also has a 1D magnon solution with energy higher than the 1D magnon mode of $\delta \bm{\tm}_+$, but lower than the bulk magnons~\cite{supplemental}.

Finally, we investigate the properties of the RM domain wall, which is both a stacking and a magnetic domain wall. The magnetic ground state is presented in Fig.~\ref{fig:magnon}(d): the magnetization of the top layer slightly tilts away from $+y$ direction in the domain wall while the magnetization of the bottom layer still roughly maintains the Walker's profile~\cite{schryer1974motion}. The Goldstone mode argument is also applicable to RM domain wall and we have verified numerically such a 1D magnon exists. Therefore, we conclude that all three types of stacking domain walls in bilayer CrI$_3$ support 1D magnon channels.

Equation \eqref{mag_energy} is oversimplified since it does not include all symmetry-allowed magnetic interactions. Especially, weak Dzyaloshinskii-Moriya interaction~\cite{dzyaloshinsky_thermodynamic_1958,moriya_anisotropic_1960} (DMI) is possible in this system. With DMI, the magnons in Fig.~\ref{fig:magnon}(a) will be generally reflection ($x \to -x$) asymmetric, which is another degree of freedom to be utilized in device designing~\cite{lan_spin-wave_2015,lan_antiferromagnetic_2017}. {Some of the 1D magnons (e.g., the 1D $\delta \bm{m}_-$ magnon in RR domain wall) do not carry net magnetic moments. For spintronics applications, a source that differentiates the top layer and the bottom layer (e.g. by proximity effects) is needed to excite these magnons~\cite{supplemental}.}

\emph{Moir\'e magnon network}.---A natural realization of stacking domain walls is by twisting the magnetic bilayer. Twisted bilayer materials create a moir\'e pattern, which is a periodic modulation of stackings. After being twisted, the structure will generally relax to lower its energy. For large twist angle, lattice relaxation can be ignored, and novel spin textures may appear~\cite{hejazi2020moir}. For small twist angles, the stable stackings will grow and form domains, while unstable stackings will shrink and eventually only appear around the domain walls.

To understand at which twist angles large domains of rhombohedral and monoclinic stackings appear and to obtain a real space pattern of these domains, we calculate the lattice relaxations of the twisted bilayer CrI$_3$ using the method introduced in Ref.~\onlinecite{carr_relaxation_2018}. We present the results of lattice relaxation for several twisting angles in the Supplemental Material~\cite{supplemental} and estimate large stacking domains emerge for twist angle smaller than $1.3^\circ$. Figure~\ref{fig:moire} shows the stacking and magnetic domain patterns in the small angle limit. Since the AFM monoclinic stacking is actually slightly energetically higher than the FM rhombohedral stacking (by about $15$ meV), the magnetic domain pattern consists of isolated AFM domains and interconnected FM domains. These domains are useful by itself. For example, using twisted bilayer CrI$_3$ as the substrate, electrons experience periodic exchange couplings in the real space and accumulate Berry phase as they move along, which may be useful to realize the topological Hall effect~\cite{bruno_topological_2004}. 

Figure~\ref{fig:moire} shows that in small angle twisted bilayer CrI$_3$, all three types of stacking domain walls appear. The interconnected stacking domain walls give rise to a magnon network, {which might be observed via inelastic tunneling spectroscopy~\cite{klein_probing_2018}}. {A similar coherent conducting network, but for electrons, can be found in twisted bilayer graphene with an out-of-plane electric field~\cite{yoo_atomic_2019,san-jose_helical_2013,huang_topologically_2018,tsim_perfect_2020,efimkin_helical_2018,xu_giant_2019,rickhaus_transport_2018}. Similarly, the 1D magnons in twisted bilayer CrI$_3$ will provide a platform for low-energy coherent spin and thermal transport. Finally, we note that beyond the continuum limit, magnons in CrI$_3$ monolayers are predicted to be topological, such that the boundary between FM and AFM regions will host chiral magnon states in the Dirac gap~\cite{aguilera_topmagnon,costa_topological_2020}.}

In summary, we propose stacking domain walls in vdW magnets can support 1D magnon channels. These channels can live on uniform magnetic ground states and are robust against external perturbations. They can be realized in naturally occurring stacking faults or through careful strain engineering~\cite{edelberg_tunable_2020}. We show that a realistic and highly tunable playground of such 1D magnons is twisted bilayer magnets with small twist angles, where their implications in spin and thermal transport phenomena are yet to be uncovered. 

\begin{acknowledgments}
We acknowledge useful discussions with Pablo Jarillo-Herrero, H\'ector Ochoa and Wenguang Zhu. This work is supported by AFOSR MURI 2D MAGIC (FA9550-19-1-0390).  The understanding of moir\'e magnon network is partially supported by DOE DE-SC0012509.  Y.G.\ and H.L.\ also acknowledge partial support from China Scholarship Council (No.~201906340219 and No.~201904910165).  Computing time is provided by BRIDGES at the Pittsburgh supercomputer center (Award No.~TG-DMR190080) under the Extreme Science and Engineering Discovery Environment (XSEDE) supported by NSF (ACI-1548562).
\end{acknowledgments}

\bibliography{main}

\begin{thebibliography}{49}%
\makeatletter
\providecommand \@ifxundefined [1]{%
 \@ifx{#1\undefined}
}%
\providecommand \@ifnum [1]{%
 \ifnum #1\expandafter \@firstoftwo
 \else \expandafter \@secondoftwo
 \fi
}%
\providecommand \@ifx [1]{%
 \ifx #1\expandafter \@firstoftwo
 \else \expandafter \@secondoftwo
 \fi
}%
\providecommand \natexlab [1]{#1}%
\providecommand \enquote  [1]{``#1''}%
\providecommand \bibnamefont  [1]{#1}%
\providecommand \bibfnamefont [1]{#1}%
\providecommand \citenamefont [1]{#1}%
\providecommand \href@noop [0]{\@secondoftwo}%
\providecommand \href [0]{\begingroup \@sanitize@url \@href}%
\providecommand \@href[1]{\@@startlink{#1}\@@href}%
\providecommand \@@href[1]{\endgroup#1\@@endlink}%
\providecommand \@sanitize@url [0]{\catcode `\\12\catcode `\$12\catcode
  `\&12\catcode `\#12\catcode `\^12\catcode `\_12\catcode `\%12\relax}%
\providecommand \@@startlink[1]{}%
\providecommand \@@endlink[0]{}%
\providecommand \url  [0]{\begingroup\@sanitize@url \@url }%
\providecommand \@url [1]{\endgroup\@href {#1}{\urlprefix }}%
\providecommand \urlprefix  [0]{URL }%
\providecommand \Eprint [0]{\href }%
\providecommand \doibase [0]{https://doi.org/}%
\providecommand \selectlanguage [0]{\@gobble}%
\providecommand \bibinfo  [0]{\@secondoftwo}%
\providecommand \bibfield  [0]{\@secondoftwo}%
\providecommand \translation [1]{[#1]}%
\providecommand \BibitemOpen [0]{}%
\providecommand \bibitemStop [0]{}%
\providecommand \bibitemNoStop [0]{.\EOS\space}%
\providecommand \EOS [0]{\spacefactor3000\relax}%
\providecommand \BibitemShut  [1]{\csname bibitem#1\endcsname}%
\let\auto@bib@innerbib\@empty
\bibitem [{\citenamefont {Alden}\ \emph {et~al.}(2013)\citenamefont {Alden},
  \citenamefont {Tsen}, \citenamefont {Huang}, \citenamefont {Hovden},
  \citenamefont {Brown}, \citenamefont {Park}, \citenamefont {Muller},\ and\
  \citenamefont {McEuen}}]{alden_strain_2013}%
  \BibitemOpen
  \bibfield  {author} {\bibinfo {author} {\bibfnamefont {J.~S.}\ \bibnamefont
  {Alden}}, \bibinfo {author} {\bibfnamefont {A.~W.}\ \bibnamefont {Tsen}},
  \bibinfo {author} {\bibfnamefont {P.~Y.}\ \bibnamefont {Huang}}, \bibinfo
  {author} {\bibfnamefont {R.}~\bibnamefont {Hovden}}, \bibinfo {author}
  {\bibfnamefont {L.}~\bibnamefont {Brown}}, \bibinfo {author} {\bibfnamefont
  {J.}~\bibnamefont {Park}}, \bibinfo {author} {\bibfnamefont {D.~A.}\
  \bibnamefont {Muller}},\ and\ \bibinfo {author} {\bibfnamefont {P.~L.}\
  \bibnamefont {McEuen}},\ }\bibfield  {title} {\bibinfo {title} {Strain
  solitons and topological defects in bilayer graphene},\ }\href
  {https://doi.org/10.1073/pnas.1309394110} {\bibfield  {journal} {\bibinfo
  {journal} {Proc. Natl. Acad. Sci.}\ }\textbf {\bibinfo {volume} {110}},\
  \bibinfo {pages} {11256} (\bibinfo {year} {2013})}\BibitemShut {NoStop}%
\bibitem [{\citenamefont {Butz}\ \emph {et~al.}(2014)\citenamefont {Butz},
  \citenamefont {Dolle}, \citenamefont {Niekiel}, \citenamefont {Weber},
  \citenamefont {Waldmann}, \citenamefont {Weber}, \citenamefont {Meyer},\ and\
  \citenamefont {Spiecker}}]{butz2014}%
  \BibitemOpen
  \bibfield  {author} {\bibinfo {author} {\bibfnamefont {B.}~\bibnamefont
  {Butz}}, \bibinfo {author} {\bibfnamefont {C.}~\bibnamefont {Dolle}},
  \bibinfo {author} {\bibfnamefont {F.}~\bibnamefont {Niekiel}}, \bibinfo
  {author} {\bibfnamefont {K.}~\bibnamefont {Weber}}, \bibinfo {author}
  {\bibfnamefont {D.}~\bibnamefont {Waldmann}}, \bibinfo {author}
  {\bibfnamefont {H.~B.}\ \bibnamefont {Weber}}, \bibinfo {author}
  {\bibfnamefont {B.}~\bibnamefont {Meyer}},\ and\ \bibinfo {author}
  {\bibfnamefont {E.}~\bibnamefont {Spiecker}},\ }\bibfield  {title} {\bibinfo
  {title} {Dislocations in bilayer graphene},\ }\href
  {https://doi.org/10.1038/nature12780} {\bibfield  {journal} {\bibinfo
  {journal} {Nature (London)}\ }\textbf {\bibinfo {volume} {505}},\ \bibinfo
  {pages} {533} (\bibinfo {year} {2014})}\BibitemShut {NoStop}%
\bibitem [{\citenamefont {Zhang}\ \emph {et~al.}(2013)\citenamefont {Zhang},
  \citenamefont {MacDonald},\ and\ \citenamefont {Mele}}]{zhang2013}%
  \BibitemOpen
  \bibfield  {author} {\bibinfo {author} {\bibfnamefont {F.}~\bibnamefont
  {Zhang}}, \bibinfo {author} {\bibfnamefont {A.~H.}\ \bibnamefont
  {MacDonald}},\ and\ \bibinfo {author} {\bibfnamefont {E.~J.}\ \bibnamefont
  {Mele}},\ }\bibfield  {title} {\bibinfo {title} {Valley {C}hern numbers and
  boundary modes in gapped bilayer graphene},\ }\href
  {https://doi.org/10.1073/pnas.1308853110} {\bibfield  {journal} {\bibinfo
  {journal} {Proc. Natl. Acad. Sci.}\ }\textbf {\bibinfo {volume} {110}},\
  \bibinfo {pages} {10546} (\bibinfo {year} {2013})}\BibitemShut {NoStop}%
\bibitem [{\citenamefont {Vaezi}\ \emph {et~al.}(2013)\citenamefont {Vaezi},
  \citenamefont {Liang}, \citenamefont {Ngai}, \citenamefont {Yang},\ and\
  \citenamefont {Kim}}]{vaezi2013}%
  \BibitemOpen
  \bibfield  {author} {\bibinfo {author} {\bibfnamefont {A.}~\bibnamefont
  {Vaezi}}, \bibinfo {author} {\bibfnamefont {Y.}~\bibnamefont {Liang}},
  \bibinfo {author} {\bibfnamefont {D.~H.}\ \bibnamefont {Ngai}}, \bibinfo
  {author} {\bibfnamefont {L.}~\bibnamefont {Yang}},\ and\ \bibinfo {author}
  {\bibfnamefont {E.-A.}\ \bibnamefont {Kim}},\ }\bibfield  {title} {\bibinfo
  {title} {Topological edge states at a tilt boundary in gated multilayer
  graphene},\ }\href {https://doi.org/10.1103/PhysRevX.3.021018} {\bibfield
  {journal} {\bibinfo  {journal} {Phys. Rev. X}\ }\textbf {\bibinfo {volume}
  {3}},\ \bibinfo {pages} {021018} (\bibinfo {year} {2013})}\BibitemShut
  {NoStop}%
\bibitem [{\citenamefont {Ju}\ \emph {et~al.}(2015)\citenamefont {Ju},
  \citenamefont {Shi}, \citenamefont {Nair}, \citenamefont {Lv}, \citenamefont
  {Jin}, \citenamefont {Velasco}, \citenamefont {Ojeda-Aristizabal},
  \citenamefont {Bechtel}, \citenamefont {Martin}, \citenamefont {Zettl},
  \citenamefont {Analytis},\ and\ \citenamefont {Wang}}]{ju_topological_2015}%
  \BibitemOpen
  \bibfield  {author} {\bibinfo {author} {\bibfnamefont {L.}~\bibnamefont
  {Ju}}, \bibinfo {author} {\bibfnamefont {Z.}~\bibnamefont {Shi}}, \bibinfo
  {author} {\bibfnamefont {N.}~\bibnamefont {Nair}}, \bibinfo {author}
  {\bibfnamefont {Y.}~\bibnamefont {Lv}}, \bibinfo {author} {\bibfnamefont
  {C.}~\bibnamefont {Jin}}, \bibinfo {author} {\bibfnamefont {J.}~\bibnamefont
  {Velasco}}, \bibinfo {author} {\bibfnamefont {C.}~\bibnamefont
  {Ojeda-Aristizabal}}, \bibinfo {author} {\bibfnamefont {H.~A.}\ \bibnamefont
  {Bechtel}}, \bibinfo {author} {\bibfnamefont {M.~C.}\ \bibnamefont {Martin}},
  \bibinfo {author} {\bibfnamefont {A.}~\bibnamefont {Zettl}}, \bibinfo
  {author} {\bibfnamefont {J.}~\bibnamefont {Analytis}},\ and\ \bibinfo
  {author} {\bibfnamefont {F.}~\bibnamefont {Wang}},\ }\bibfield  {title}
  {\bibinfo {title} {Topological valley transport at bilayer graphene domain
  walls},\ }\href {https://doi.org/10.1038/nature14364} {\bibfield  {journal}
  {\bibinfo  {journal} {Nature (London)}\ }\textbf {\bibinfo {volume} {520}},\
  \bibinfo {pages} {650} (\bibinfo {year} {2015})}\BibitemShut {NoStop}%
\bibitem [{\citenamefont {Yin}\ \emph {et~al.}(2016)\citenamefont {Yin},
  \citenamefont {Jiang}, \citenamefont {Qiao},\ and\ \citenamefont
  {He}}]{yin_direct_2016}%
  \BibitemOpen
  \bibfield  {author} {\bibinfo {author} {\bibfnamefont {L.-J.}\ \bibnamefont
  {Yin}}, \bibinfo {author} {\bibfnamefont {H.}~\bibnamefont {Jiang}}, \bibinfo
  {author} {\bibfnamefont {J.-B.}\ \bibnamefont {Qiao}},\ and\ \bibinfo
  {author} {\bibfnamefont {L.}~\bibnamefont {He}},\ }\bibfield  {title}
  {\bibinfo {title} {Direct imaging of topological edge states at a bilayer
  graphene domain wall},\ }\href {https://doi.org/10.1038/ncomms11760}
  {\bibfield  {journal} {\bibinfo  {journal} {Nat. Commun.}\ }\textbf {\bibinfo
  {volume} {7}},\ \bibinfo {pages} {11760} (\bibinfo {year}
  {2016})}\BibitemShut {NoStop}%
\bibitem [{\citenamefont {Edelberg}\ \emph {et~al.}(2020)\citenamefont
  {Edelberg}, \citenamefont {Kumar}, \citenamefont {Shenoy}, \citenamefont
  {Ochoa},\ and\ \citenamefont {Pasupathy}}]{edelberg_tunable_2020}%
  \BibitemOpen
  \bibfield  {author} {\bibinfo {author} {\bibfnamefont {D.}~\bibnamefont
  {Edelberg}}, \bibinfo {author} {\bibfnamefont {H.}~\bibnamefont {Kumar}},
  \bibinfo {author} {\bibfnamefont {V.}~\bibnamefont {Shenoy}}, \bibinfo
  {author} {\bibfnamefont {H.}~\bibnamefont {Ochoa}},\ and\ \bibinfo {author}
  {\bibfnamefont {A.~N.}\ \bibnamefont {Pasupathy}},\ }\bibfield  {title}
  {\bibinfo {title} {Tunable strain soliton networks confine electrons in van
  der {Waals} materials},\ }\href {https://doi.org/10.1038/s41567-020-0953-2}
  {\bibfield  {journal} {\bibinfo  {journal} {Nat. Phys.}\ ,\ \bibinfo {pages}
  {1}} (\bibinfo {year} {2020})}\BibitemShut {NoStop}%
\bibitem [{\citenamefont {Sivadas}\ \emph {et~al.}(2018)\citenamefont
  {Sivadas}, \citenamefont {Okamoto}, \citenamefont {Xu}, \citenamefont
  {Fennie},\ and\ \citenamefont {Xiao}}]{sivadas_stacking-dependent_2018}%
  \BibitemOpen
  \bibfield  {author} {\bibinfo {author} {\bibfnamefont {N.}~\bibnamefont
  {Sivadas}}, \bibinfo {author} {\bibfnamefont {S.}~\bibnamefont {Okamoto}},
  \bibinfo {author} {\bibfnamefont {X.}~\bibnamefont {Xu}}, \bibinfo {author}
  {\bibfnamefont {C.~J.}\ \bibnamefont {Fennie}},\ and\ \bibinfo {author}
  {\bibfnamefont {D.}~\bibnamefont {Xiao}},\ }\bibfield  {title} {\bibinfo
  {title} {Stacking-dependent magnetism in bilayer {CrI$_3$}},\ }\href
  {https://doi.org/10.1021/acs.nanolett.8b03321} {\bibfield  {journal}
  {\bibinfo  {journal} {Nano Lett.}\ }\textbf {\bibinfo {volume} {18}},\
  \bibinfo {pages} {7658} (\bibinfo {year} {2018})}\BibitemShut {NoStop}%
\bibitem [{\citenamefont {Wang}\ \emph {et~al.}(2018)\citenamefont {Wang},
  \citenamefont {Guti{\'e}rrez-Lezama}, \citenamefont {Ubrig}, \citenamefont
  {Kroner}, \citenamefont {Gibertini}, \citenamefont {Taniguchi}, \citenamefont
  {Watanabe}, \citenamefont {Imamo{\u g}lu}, \citenamefont {Giannini},\ and\
  \citenamefont {Morpurgo}}]{wang2018}%
  \BibitemOpen
  \bibfield  {author} {\bibinfo {author} {\bibfnamefont {Z.}~\bibnamefont
  {Wang}}, \bibinfo {author} {\bibfnamefont {I.}~\bibnamefont
  {Guti{\'e}rrez-Lezama}}, \bibinfo {author} {\bibfnamefont {N.}~\bibnamefont
  {Ubrig}}, \bibinfo {author} {\bibfnamefont {M.}~\bibnamefont {Kroner}},
  \bibinfo {author} {\bibfnamefont {M.}~\bibnamefont {Gibertini}}, \bibinfo
  {author} {\bibfnamefont {T.}~\bibnamefont {Taniguchi}}, \bibinfo {author}
  {\bibfnamefont {K.}~\bibnamefont {Watanabe}}, \bibinfo {author}
  {\bibfnamefont {A.}~\bibnamefont {Imamo{\u g}lu}}, \bibinfo {author}
  {\bibfnamefont {E.}~\bibnamefont {Giannini}},\ and\ \bibinfo {author}
  {\bibfnamefont {A.~F.}\ \bibnamefont {Morpurgo}},\ }\bibfield  {title}
  {\bibinfo {title} {Very large tunneling magnetoresistance in layered magnetic
  semiconductor {CrI$_3$}},\ }\href
  {https://doi.org/10.1038/s41467-018-04953-8} {\bibfield  {journal} {\bibinfo
  {journal} {Nat. Commun.}\ }\textbf {\bibinfo {volume} {9}},\ \bibinfo {pages}
  {2516} (\bibinfo {year} {2018})}\BibitemShut {NoStop}%
\bibitem [{\citenamefont {Jiang}\ \emph {et~al.}(2019)\citenamefont {Jiang},
  \citenamefont {Wang}, \citenamefont {Chen}, \citenamefont {Zhong},
  \citenamefont {Yuan}, \citenamefont {Lu},\ and\ \citenamefont
  {Ji}}]{jiang_stacking_2019}%
  \BibitemOpen
  \bibfield  {author} {\bibinfo {author} {\bibfnamefont {P.}~\bibnamefont
  {Jiang}}, \bibinfo {author} {\bibfnamefont {C.}~\bibnamefont {Wang}},
  \bibinfo {author} {\bibfnamefont {D.}~\bibnamefont {Chen}}, \bibinfo {author}
  {\bibfnamefont {Z.}~\bibnamefont {Zhong}}, \bibinfo {author} {\bibfnamefont
  {Z.}~\bibnamefont {Yuan}}, \bibinfo {author} {\bibfnamefont {Z.-Y.}\
  \bibnamefont {Lu}},\ and\ \bibinfo {author} {\bibfnamefont {W.}~\bibnamefont
  {Ji}},\ }\bibfield  {title} {\bibinfo {title} {Stacking tunable interlayer
  magnetism in bilayer {CrI$_3$}},\ }\href
  {https://doi.org/10.1103/PhysRevB.99.144401} {\bibfield  {journal} {\bibinfo
  {journal} {Phys. Rev. B}\ }\textbf {\bibinfo {volume} {99}},\ \bibinfo
  {pages} {144401} (\bibinfo {year} {2019})}\BibitemShut {NoStop}%
\bibitem [{\citenamefont {Soriano}\ \emph {et~al.}(2019)\citenamefont
  {Soriano}, \citenamefont {Cardoso},\ and\ \citenamefont
  {Fern{\'a}ndez-Rossier}}]{soriano2019}%
  \BibitemOpen
  \bibfield  {author} {\bibinfo {author} {\bibfnamefont {D.}~\bibnamefont
  {Soriano}}, \bibinfo {author} {\bibfnamefont {C.}~\bibnamefont {Cardoso}},\
  and\ \bibinfo {author} {\bibfnamefont {J.}~\bibnamefont
  {Fern{\'a}ndez-Rossier}},\ }\bibfield  {title} {\bibinfo {title} {Interplay
  between interlayer exchange and stacking in {CrI$_3$} bilayers},\ }\href
  {https://doi.org/https://doi.org/10.1016/j.ssc.2019.113662} {\bibfield
  {journal} {\bibinfo  {journal} {Solid State Commun.}\ }\textbf {\bibinfo
  {volume} {299}},\ \bibinfo {pages} {113662} (\bibinfo {year}
  {2019})}\BibitemShut {NoStop}%
\bibitem [{\citenamefont {Jang}\ \emph {et~al.}(2019)\citenamefont {Jang},
  \citenamefont {Jeong}, \citenamefont {Yoon}, \citenamefont {Ryee},\ and\
  \citenamefont {Han}}]{jang2019}%
  \BibitemOpen
  \bibfield  {author} {\bibinfo {author} {\bibfnamefont {S.~W.}\ \bibnamefont
  {Jang}}, \bibinfo {author} {\bibfnamefont {M.~Y.}\ \bibnamefont {Jeong}},
  \bibinfo {author} {\bibfnamefont {H.}~\bibnamefont {Yoon}}, \bibinfo {author}
  {\bibfnamefont {S.}~\bibnamefont {Ryee}},\ and\ \bibinfo {author}
  {\bibfnamefont {M.~J.}\ \bibnamefont {Han}},\ }\bibfield  {title} {\bibinfo
  {title} {Microscopic understanding of magnetic interactions in bilayer
  {CrI}$_3$},\ }\href {https://doi.org/10.1103/PhysRevMaterials.3.031001}
  {\bibfield  {journal} {\bibinfo  {journal} {Phys. Rev. Mater.}\ }\textbf
  {\bibinfo {volume} {3}},\ \bibinfo {pages} {031001} (\bibinfo {year}
  {2019})}\BibitemShut {NoStop}%
\bibitem [{\citenamefont {Chen}\ \emph {et~al.}(2019)\citenamefont {Chen},
  \citenamefont {Sun}, \citenamefont {Wang}, \citenamefont {Gu}, \citenamefont
  {Xu}, \citenamefont {Wu},\ and\ \citenamefont {Gao}}]{chen2019direct}%
  \BibitemOpen
  \bibfield  {author} {\bibinfo {author} {\bibfnamefont {W.}~\bibnamefont
  {Chen}}, \bibinfo {author} {\bibfnamefont {Z.}~\bibnamefont {Sun}}, \bibinfo
  {author} {\bibfnamefont {Z.}~\bibnamefont {Wang}}, \bibinfo {author}
  {\bibfnamefont {L.}~\bibnamefont {Gu}}, \bibinfo {author} {\bibfnamefont
  {X.}~\bibnamefont {Xu}}, \bibinfo {author} {\bibfnamefont {S.}~\bibnamefont
  {Wu}},\ and\ \bibinfo {author} {\bibfnamefont {C.}~\bibnamefont {Gao}},\
  }\bibfield  {title} {\bibinfo {title} {Direct observation of van der waals
  stacking--dependent interlayer magnetism},\ }\href@noop {} {\bibfield
  {journal} {\bibinfo  {journal} {Science}\ }\textbf {\bibinfo {volume}
  {366}},\ \bibinfo {pages} {983} (\bibinfo {year} {2019})}\BibitemShut
  {NoStop}%
\bibitem [{\citenamefont {Winter}(1961)}]{winter_bloch_1961}%
  \BibitemOpen
  \bibfield  {author} {\bibinfo {author} {\bibfnamefont {J.~M.}\ \bibnamefont
  {Winter}},\ }\bibfield  {title} {\bibinfo {title} {Bloch {Wall} {Excitation}.
  {Application} to {Nuclear} {Resonance} in a {Bloch} {Wall}},\ }\href
  {https://doi.org/10.1103/PhysRev.124.452} {\bibfield  {journal} {\bibinfo
  {journal} {Phys. Rev.}\ }\textbf {\bibinfo {volume} {124}},\ \bibinfo {pages}
  {452} (\bibinfo {year} {1961})}\BibitemShut {NoStop}%
\bibitem [{\citenamefont {Thiele}(1976)}]{thiele_excitation_1976}%
  \BibitemOpen
  \bibfield  {author} {\bibinfo {author} {\bibfnamefont {A.~A.}\ \bibnamefont
  {Thiele}},\ }\bibfield  {title} {\bibinfo {title} {Excitation spectrum of a
  magnetic domain wall containing {Bloch} lines},\ }\href
  {https://doi.org/10.1103/PhysRevB.14.3130} {\bibfield  {journal} {\bibinfo
  {journal} {Phys. Rev. B}\ }\textbf {\bibinfo {volume} {14}},\ \bibinfo
  {pages} {3130} (\bibinfo {year} {1976})}\BibitemShut {NoStop}%
\bibitem [{\citenamefont {Ferrer}\ \emph {et~al.}(2003)\citenamefont {Ferrer},
  \citenamefont {Farinas},\ and\ \citenamefont
  {Caldeira}}]{ferrer_one-dimensional_2003}%
  \BibitemOpen
  \bibfield  {author} {\bibinfo {author} {\bibfnamefont {A.~V.}\ \bibnamefont
  {Ferrer}}, \bibinfo {author} {\bibfnamefont {P.~F.}\ \bibnamefont
  {Farinas}},\ and\ \bibinfo {author} {\bibfnamefont {A.~O.}\ \bibnamefont
  {Caldeira}},\ }\bibfield  {title} {\bibinfo {title} {One-{Dimensional}
  {Gapless} {Magnons} in a {Single} {Anisotropic} {Ferromagnetic}
  {Nanolayer}},\ }\href
  {https://link.aps.org/doi/10.1103/PhysRevLett.91.226803} {\bibfield
  {journal} {\bibinfo  {journal} {Phys. Rev. Lett.}\ }\textbf {\bibinfo
  {volume} {91}},\ \bibinfo {pages} {226803} (\bibinfo {year}
  {2003})}\BibitemShut {NoStop}%
\bibitem [{\citenamefont {Garcia-Sanchez}\ \emph {et~al.}(2015)\citenamefont
  {Garcia-Sanchez}, \citenamefont {Borys}, \citenamefont {Soucaille},
  \citenamefont {Adam}, \citenamefont {Stamps},\ and\ \citenamefont
  {Kim}}]{Sanchez_narrow_2015}%
  \BibitemOpen
  \bibfield  {author} {\bibinfo {author} {\bibfnamefont {F.}~\bibnamefont
  {Garcia-Sanchez}}, \bibinfo {author} {\bibfnamefont {P.}~\bibnamefont
  {Borys}}, \bibinfo {author} {\bibfnamefont {R.}~\bibnamefont {Soucaille}},
  \bibinfo {author} {\bibfnamefont {J.-P.}\ \bibnamefont {Adam}}, \bibinfo
  {author} {\bibfnamefont {R.~L.}\ \bibnamefont {Stamps}},\ and\ \bibinfo
  {author} {\bibfnamefont {J.-V.}\ \bibnamefont {Kim}},\ }\bibfield  {title}
  {\bibinfo {title} {Narrow magnonic waveguides based on domain walls},\ }\href
  {https://doi.org/10.1103/PhysRevLett.114.247206} {\bibfield  {journal}
  {\bibinfo  {journal} {Phys. Rev. Lett.}\ }\textbf {\bibinfo {volume} {114}},\
  \bibinfo {pages} {247206} (\bibinfo {year} {2015})}\BibitemShut {NoStop}%
\bibitem [{\citenamefont {Lan}\ \emph {et~al.}(2015)\citenamefont {Lan},
  \citenamefont {Yu}, \citenamefont {Wu},\ and\ \citenamefont
  {Xiao}}]{lan_spin-wave_2015}%
  \BibitemOpen
  \bibfield  {author} {\bibinfo {author} {\bibfnamefont {J.}~\bibnamefont
  {Lan}}, \bibinfo {author} {\bibfnamefont {W.}~\bibnamefont {Yu}}, \bibinfo
  {author} {\bibfnamefont {R.}~\bibnamefont {Wu}},\ and\ \bibinfo {author}
  {\bibfnamefont {J.}~\bibnamefont {Xiao}},\ }\bibfield  {title} {\bibinfo
  {title} {Spin-{Wave} {Diode}},\ }\href
  {https://doi.org/10.1103/PhysRevX.5.041049} {\bibfield  {journal} {\bibinfo
  {journal} {Phys. Rev. X}\ }\textbf {\bibinfo {volume} {5}},\ \bibinfo {pages}
  {041049} (\bibinfo {year} {2015})}\BibitemShut {NoStop}%
\bibitem [{\citenamefont {Hejazi}\ \emph {et~al.}(2020)\citenamefont {Hejazi},
  \citenamefont {Luo},\ and\ \citenamefont {Balents}}]{hejazi2020moir}%
  \BibitemOpen
  \bibfield  {author} {\bibinfo {author} {\bibfnamefont {K.}~\bibnamefont
  {Hejazi}}, \bibinfo {author} {\bibfnamefont {Z.-X.}\ \bibnamefont {Luo}},\
  and\ \bibinfo {author} {\bibfnamefont {L.}~\bibnamefont {Balents}},\
  }\bibfield  {title} {\bibinfo {title} {Noncollinear phases in moir\'e
  magnets},\ }\href {https://doi.org/10.1073/pnas.2000347117} {\bibfield
  {journal} {\bibinfo  {journal} {Proc. Natl. Acad. Sci. U.S.A.}\ }\textbf
  {\bibinfo {volume} {117}},\ \bibinfo {pages} {10721} (\bibinfo {year}
  {2020})}\BibitemShut {NoStop}%
\bibitem [{\citenamefont {Li}\ and\ \citenamefont
  {Cheng}(2020)}]{li_moire_2020}%
  \BibitemOpen
  \bibfield  {author} {\bibinfo {author} {\bibfnamefont {Y.-H.}\ \bibnamefont
  {Li}}\ and\ \bibinfo {author} {\bibfnamefont {R.}~\bibnamefont {Cheng}},\
  }\bibfield  {title} {\bibinfo {title} {Moir\'e magnons in twisted bilayer
  magnets with collinear order},\ }\href
  {https://doi.org/10.1103/PhysRevB.102.094404} {\bibfield  {journal} {\bibinfo
   {journal} {Phys. Rev. B}\ }\textbf {\bibinfo {volume} {102}},\ \bibinfo
  {pages} {094404} (\bibinfo {year} {2020})}\BibitemShut {NoStop}%
\bibitem [{\citenamefont {Ghader}(2020)}]{ghader_magnon_2020}%
  \BibitemOpen
  \bibfield  {author} {\bibinfo {author} {\bibfnamefont {D.}~\bibnamefont
  {Ghader}},\ }\bibfield  {title} {\bibinfo {title} {Magnon magic angles and
  tunable {Hall} conductivity in {2D} twisted ferromagnetic bilayers},\
  }\href@noop {} {\bibfield  {journal} {\bibinfo  {journal} {Sci. Rep.}\
  }\textbf {\bibinfo {volume} {10}},\ \bibinfo {pages} {1} (\bibinfo {year}
  {2020})}\BibitemShut {NoStop}%
\bibitem [{\citenamefont {Huang}\ \emph {et~al.}(2017)\citenamefont {Huang},
  \citenamefont {Clark}, \citenamefont {Navarro-Moratalla}, \citenamefont
  {Klein}, \citenamefont {Cheng}, \citenamefont {Seyler}, \citenamefont
  {Zhong}, \citenamefont {Schmidgall}, \citenamefont {McGuire}, \citenamefont
  {Cobden}, \citenamefont {Yao}, \citenamefont {Xiao}, \citenamefont
  {Jarillo-Herrero},\ and\ \citenamefont {Xu}}]{huang_layer-dependent_2017}%
  \BibitemOpen
  \bibfield  {author} {\bibinfo {author} {\bibfnamefont {B.}~\bibnamefont
  {Huang}}, \bibinfo {author} {\bibfnamefont {G.}~\bibnamefont {Clark}},
  \bibinfo {author} {\bibfnamefont {E.}~\bibnamefont {Navarro-Moratalla}},
  \bibinfo {author} {\bibfnamefont {D.~R.}\ \bibnamefont {Klein}}, \bibinfo
  {author} {\bibfnamefont {R.}~\bibnamefont {Cheng}}, \bibinfo {author}
  {\bibfnamefont {K.~L.}\ \bibnamefont {Seyler}}, \bibinfo {author}
  {\bibfnamefont {D.}~\bibnamefont {Zhong}}, \bibinfo {author} {\bibfnamefont
  {E.}~\bibnamefont {Schmidgall}}, \bibinfo {author} {\bibfnamefont {M.~A.}\
  \bibnamefont {McGuire}}, \bibinfo {author} {\bibfnamefont {D.~H.}\
  \bibnamefont {Cobden}}, \bibinfo {author} {\bibfnamefont {W.}~\bibnamefont
  {Yao}}, \bibinfo {author} {\bibfnamefont {D.}~\bibnamefont {Xiao}}, \bibinfo
  {author} {\bibfnamefont {P.}~\bibnamefont {Jarillo-Herrero}},\ and\ \bibinfo
  {author} {\bibfnamefont {X.}~\bibnamefont {Xu}},\ }\bibfield  {title}
  {\bibinfo {title} {Layer-dependent ferromagnetism in a van der {Waals}
  crystal down to the monolayer limit},\ }\href
  {https://doi.org/10.1038/nature22391} {\bibfield  {journal} {\bibinfo
  {journal} {Nature (London)}\ }\textbf {\bibinfo {volume} {546}},\ \bibinfo
  {pages} {270} (\bibinfo {year} {2017})}\BibitemShut {NoStop}%
\bibitem [{sup()}]{supplemental}%
  \BibitemOpen
  \href@noop {} {}\bibinfo {note} {Supplemental Material presenting details
  about DFT methods, lattice relaxation of the CrI$_3$ moir\'e structures, the
  magnetic property of the MM domain wall, the dispersion of some 1D magnons
  and a discussion of how to excite 1D magnons. The Supplemental Material cites
  Refs.
  \onlinecite{vasp1,vasp2,blochl_projector_1994,PBEsol,carr_relaxation_2018,bezanson2017julia,rackauckas2017differentialequations,gargiulo_structural_2017,DFT_D3,zhang_spin_2012}}\BibitemShut
  {NoStop}%
\bibitem [{\citenamefont {Cenker}\ \emph {et~al.}(2020)\citenamefont {Cenker},
  \citenamefont {Huang}, \citenamefont {Suri}, \citenamefont {Thijssen},
  \citenamefont {Miller}, \citenamefont {Song}, \citenamefont {Taniguchi},
  \citenamefont {Watanabe}, \citenamefont {McGuire}, \citenamefont {Xiao},\
  and\ \citenamefont {Xu}}]{cenker_direct_2020}%
  \BibitemOpen
  \bibfield  {author} {\bibinfo {author} {\bibfnamefont {J.}~\bibnamefont
  {Cenker}}, \bibinfo {author} {\bibfnamefont {B.}~\bibnamefont {Huang}},
  \bibinfo {author} {\bibfnamefont {N.}~\bibnamefont {Suri}}, \bibinfo {author}
  {\bibfnamefont {P.}~\bibnamefont {Thijssen}}, \bibinfo {author}
  {\bibfnamefont {A.}~\bibnamefont {Miller}}, \bibinfo {author} {\bibfnamefont
  {T.}~\bibnamefont {Song}}, \bibinfo {author} {\bibfnamefont {T.}~\bibnamefont
  {Taniguchi}}, \bibinfo {author} {\bibfnamefont {K.}~\bibnamefont {Watanabe}},
  \bibinfo {author} {\bibfnamefont {M.~A.}\ \bibnamefont {McGuire}}, \bibinfo
  {author} {\bibfnamefont {D.}~\bibnamefont {Xiao}},\ and\ \bibinfo {author}
  {\bibfnamefont {X.}~\bibnamefont {Xu}},\ }\bibfield  {title} {\bibinfo
  {title} {Direct observation of two-dimensional magnons in atomically thin
  {CrI}$_3$},\ }\href {https://doi.org/10.1038/s41567-020-0999-1} {\bibfield
  {journal} {\bibinfo  {journal} {Nat. Phys.}\ ,\ \bibinfo {pages} {1}}
  (\bibinfo {year} {2020})}\BibitemShut {NoStop}%
\bibitem [{\citenamefont {Schryer}\ and\ \citenamefont
  {Walker}(1974)}]{schryer1974motion}%
  \BibitemOpen
  \bibfield  {author} {\bibinfo {author} {\bibfnamefont {N.~L.}\ \bibnamefont
  {Schryer}}\ and\ \bibinfo {author} {\bibfnamefont {L.~R.}\ \bibnamefont
  {Walker}},\ }\bibfield  {title} {\bibinfo {title} {The motion of 180 domain
  walls in uniform dc magnetic fields},\ }\href@noop {} {\bibfield  {journal}
  {\bibinfo  {journal} {J. Appl. Phys.}\ }\textbf {\bibinfo {volume} {45}},\
  \bibinfo {pages} {5406} (\bibinfo {year} {1974})}\BibitemShut {NoStop}%
\bibitem [{\citenamefont
  {Dzyaloshinsky}(1958)}]{dzyaloshinsky_thermodynamic_1958}%
  \BibitemOpen
  \bibfield  {author} {\bibinfo {author} {\bibfnamefont {I.}~\bibnamefont
  {Dzyaloshinsky}},\ }\bibfield  {title} {\bibinfo {title} {A thermodynamic
  theory of ``weak'' ferromagnetism of antiferromagnetics},\ }\href
  {https://doi.org/10.1016/0022-3697(58)90076-3} {\bibfield  {journal}
  {\bibinfo  {journal} {J Phys. Chem. Solids}\ }\textbf {\bibinfo {volume}
  {4}},\ \bibinfo {pages} {241} (\bibinfo {year} {1958})}\BibitemShut {NoStop}%
\bibitem [{\citenamefont {Moriya}(1960)}]{moriya_anisotropic_1960}%
  \BibitemOpen
  \bibfield  {author} {\bibinfo {author} {\bibfnamefont {T.}~\bibnamefont
  {Moriya}},\ }\bibfield  {title} {\bibinfo {title} {Anisotropic
  {Superexchange} {Interaction} and {Weak} {Ferromagnetism}},\ }\href
  {https://doi.org/10.1103/PhysRev.120.91} {\bibfield  {journal} {\bibinfo
  {journal} {Phys. Rev.}\ }\textbf {\bibinfo {volume} {120}},\ \bibinfo {pages}
  {91} (\bibinfo {year} {1960})}\BibitemShut {NoStop}%
\bibitem [{\citenamefont {Lan}\ \emph {et~al.}(2017)\citenamefont {Lan},
  \citenamefont {Yu},\ and\ \citenamefont {Xiao}}]{lan_antiferromagnetic_2017}%
  \BibitemOpen
  \bibfield  {author} {\bibinfo {author} {\bibfnamefont {J.}~\bibnamefont
  {Lan}}, \bibinfo {author} {\bibfnamefont {W.}~\bibnamefont {Yu}},\ and\
  \bibinfo {author} {\bibfnamefont {J.}~\bibnamefont {Xiao}},\ }\bibfield
  {title} {\bibinfo {title} {Antiferromagnetic domain wall as spin wave
  polarizer and retarder},\ }\href {https://doi.org/10.1038/s41467-017-00265-5}
  {\bibfield  {journal} {\bibinfo  {journal} {Nat. Commun.}\ }\textbf {\bibinfo
  {volume} {8}},\ \bibinfo {pages} {178} (\bibinfo {year} {2017})}\BibitemShut
  {NoStop}%
\bibitem [{\citenamefont {Carr}\ \emph {et~al.}(2018)\citenamefont {Carr},
  \citenamefont {Massatt}, \citenamefont {Torrisi}, \citenamefont {Cazeaux},
  \citenamefont {Luskin},\ and\ \citenamefont
  {Kaxiras}}]{carr_relaxation_2018}%
  \BibitemOpen
  \bibfield  {author} {\bibinfo {author} {\bibfnamefont {S.}~\bibnamefont
  {Carr}}, \bibinfo {author} {\bibfnamefont {D.}~\bibnamefont {Massatt}},
  \bibinfo {author} {\bibfnamefont {S.~B.}\ \bibnamefont {Torrisi}}, \bibinfo
  {author} {\bibfnamefont {P.}~\bibnamefont {Cazeaux}}, \bibinfo {author}
  {\bibfnamefont {M.}~\bibnamefont {Luskin}},\ and\ \bibinfo {author}
  {\bibfnamefont {E.}~\bibnamefont {Kaxiras}},\ }\bibfield  {title} {\bibinfo
  {title} {Relaxation and domain formation in incommensurate two-dimensional
  heterostructures},\ }\href {https://doi.org/10.1103/PhysRevB.98.224102}
  {\bibfield  {journal} {\bibinfo  {journal} {Phys. Rev. B}\ }\textbf {\bibinfo
  {volume} {98}},\ \bibinfo {pages} {224102} (\bibinfo {year}
  {2018})}\BibitemShut {NoStop}%
\bibitem [{\citenamefont {Bruno}\ \emph {et~al.}(2004)\citenamefont {Bruno},
  \citenamefont {Dugaev},\ and\ \citenamefont
  {Taillefumier}}]{bruno_topological_2004}%
  \BibitemOpen
  \bibfield  {author} {\bibinfo {author} {\bibfnamefont {P.}~\bibnamefont
  {Bruno}}, \bibinfo {author} {\bibfnamefont {V.~K.}\ \bibnamefont {Dugaev}},\
  and\ \bibinfo {author} {\bibfnamefont {M.}~\bibnamefont {Taillefumier}},\
  }\bibfield  {title} {\bibinfo {title} {Topological hall effect and {Berry}
  phase in magnetic nanostructures},\ }\href
  {https://doi.org/10.1103/PhysRevLett.93.096806} {\bibfield  {journal}
  {\bibinfo  {journal} {Phys. Rev. Lett.}\ }\textbf {\bibinfo {volume} {93}},\
  \bibinfo {pages} {096806} (\bibinfo {year} {2004})}\BibitemShut {NoStop}%
\bibitem [{\citenamefont {Klein}\ \emph {et~al.}(2018)\citenamefont {Klein},
  \citenamefont {MacNeill}, \citenamefont {Lado}, \citenamefont {Soriano},
  \citenamefont {Navarro-Moratalla}, \citenamefont {Watanabe}, \citenamefont
  {Taniguchi}, \citenamefont {Manni}, \citenamefont {Canfield}, \citenamefont
  {Fernández-Rossier},\ and\ \citenamefont
  {Jarillo-Herrero}}]{klein_probing_2018}%
  \BibitemOpen
  \bibfield  {author} {\bibinfo {author} {\bibfnamefont {D.~R.}\ \bibnamefont
  {Klein}}, \bibinfo {author} {\bibfnamefont {D.}~\bibnamefont {MacNeill}},
  \bibinfo {author} {\bibfnamefont {J.~L.}\ \bibnamefont {Lado}}, \bibinfo
  {author} {\bibfnamefont {D.}~\bibnamefont {Soriano}}, \bibinfo {author}
  {\bibfnamefont {E.}~\bibnamefont {Navarro-Moratalla}}, \bibinfo {author}
  {\bibfnamefont {K.}~\bibnamefont {Watanabe}}, \bibinfo {author}
  {\bibfnamefont {T.}~\bibnamefont {Taniguchi}}, \bibinfo {author}
  {\bibfnamefont {S.}~\bibnamefont {Manni}}, \bibinfo {author} {\bibfnamefont
  {P.}~\bibnamefont {Canfield}}, \bibinfo {author} {\bibfnamefont
  {J.}~\bibnamefont {Fernández-Rossier}},\ and\ \bibinfo {author}
  {\bibfnamefont {P.}~\bibnamefont {Jarillo-Herrero}},\ }\bibfield  {title}
  {\bibinfo {title} {Probing magnetism in {2D} van der {Waals} crystalline
  insulators via electron tunneling},\ }\href
  {https://doi.org/10.1126/science.aar3617} {\bibfield  {journal} {\bibinfo
  {journal} {Science}\ }\textbf {\bibinfo {volume} {360}},\ \bibinfo {pages}
  {1218} (\bibinfo {year} {2018})}\BibitemShut {NoStop}%
\bibitem [{\citenamefont {Yoo}\ \emph {et~al.}(2019)\citenamefont {Yoo},
  \citenamefont {Engelke}, \citenamefont {Carr}, \citenamefont {Fang},
  \citenamefont {Zhang}, \citenamefont {Cazeaux}, \citenamefont {Sung},
  \citenamefont {Hovden}, \citenamefont {Tsen}, \citenamefont {Taniguchi},
  \citenamefont {Watanabe}, \citenamefont {Yi}, \citenamefont {Kim},
  \citenamefont {Luskin}, \citenamefont {Tadmor}, \citenamefont {Kaxiras},\
  and\ \citenamefont {Kim}}]{yoo_atomic_2019}%
  \BibitemOpen
  \bibfield  {author} {\bibinfo {author} {\bibfnamefont {H.}~\bibnamefont
  {Yoo}}, \bibinfo {author} {\bibfnamefont {R.}~\bibnamefont {Engelke}},
  \bibinfo {author} {\bibfnamefont {S.}~\bibnamefont {Carr}}, \bibinfo {author}
  {\bibfnamefont {S.}~\bibnamefont {Fang}}, \bibinfo {author} {\bibfnamefont
  {K.}~\bibnamefont {Zhang}}, \bibinfo {author} {\bibfnamefont
  {P.}~\bibnamefont {Cazeaux}}, \bibinfo {author} {\bibfnamefont {S.~H.}\
  \bibnamefont {Sung}}, \bibinfo {author} {\bibfnamefont {R.}~\bibnamefont
  {Hovden}}, \bibinfo {author} {\bibfnamefont {A.~W.}\ \bibnamefont {Tsen}},
  \bibinfo {author} {\bibfnamefont {T.}~\bibnamefont {Taniguchi}}, \bibinfo
  {author} {\bibfnamefont {K.}~\bibnamefont {Watanabe}}, \bibinfo {author}
  {\bibfnamefont {G.-C.}\ \bibnamefont {Yi}}, \bibinfo {author} {\bibfnamefont
  {M.}~\bibnamefont {Kim}}, \bibinfo {author} {\bibfnamefont {M.}~\bibnamefont
  {Luskin}}, \bibinfo {author} {\bibfnamefont {E.~B.}\ \bibnamefont {Tadmor}},
  \bibinfo {author} {\bibfnamefont {E.}~\bibnamefont {Kaxiras}},\ and\ \bibinfo
  {author} {\bibfnamefont {P.}~\bibnamefont {Kim}},\ }\bibfield  {title}
  {\bibinfo {title} {Atomic and electronic reconstruction at the van der
  {Waals} interface in twisted bilayer graphene},\ }\href
  {https://doi.org/10.1038/s41563-019-0346-z} {\bibfield  {journal} {\bibinfo
  {journal} {Nat. Mater.}\ }\textbf {\bibinfo {volume} {18}},\ \bibinfo {pages}
  {448} (\bibinfo {year} {2019})}\BibitemShut {NoStop}%
\bibitem [{\citenamefont {San-Jose}\ and\ \citenamefont
  {Prada}(2013)}]{san-jose_helical_2013}%
  \BibitemOpen
  \bibfield  {author} {\bibinfo {author} {\bibfnamefont {P.}~\bibnamefont
  {San-Jose}}\ and\ \bibinfo {author} {\bibfnamefont {E.}~\bibnamefont
  {Prada}},\ }\bibfield  {title} {\bibinfo {title} {Helical networks in twisted
  bilayer graphene under interlayer bias},\ }\href
  {https://doi.org/10.1103/PhysRevB.88.121408} {\bibfield  {journal} {\bibinfo
  {journal} {Phys. Rev. B}\ }\textbf {\bibinfo {volume} {88}},\ \bibinfo
  {pages} {121408} (\bibinfo {year} {2013})}\BibitemShut {NoStop}%
\bibitem [{\citenamefont {Huang}\ \emph {et~al.}(2018)\citenamefont {Huang},
  \citenamefont {Kim}, \citenamefont {Efimkin}, \citenamefont {Lovorn},
  \citenamefont {Taniguchi}, \citenamefont {Watanabe}, \citenamefont
  {MacDonald}, \citenamefont {Tutuc},\ and\ \citenamefont
  {LeRoy}}]{huang_topologically_2018}%
  \BibitemOpen
  \bibfield  {author} {\bibinfo {author} {\bibfnamefont {S.}~\bibnamefont
  {Huang}}, \bibinfo {author} {\bibfnamefont {K.}~\bibnamefont {Kim}}, \bibinfo
  {author} {\bibfnamefont {D.~K.}\ \bibnamefont {Efimkin}}, \bibinfo {author}
  {\bibfnamefont {T.}~\bibnamefont {Lovorn}}, \bibinfo {author} {\bibfnamefont
  {T.}~\bibnamefont {Taniguchi}}, \bibinfo {author} {\bibfnamefont
  {K.}~\bibnamefont {Watanabe}}, \bibinfo {author} {\bibfnamefont {A.~H.}\
  \bibnamefont {MacDonald}}, \bibinfo {author} {\bibfnamefont {E.}~\bibnamefont
  {Tutuc}},\ and\ \bibinfo {author} {\bibfnamefont {B.~J.}\ \bibnamefont
  {LeRoy}},\ }\bibfield  {title} {\bibinfo {title} {Topologically {Protected}
  {Helical} {States} in {Minimally} {Twisted} {Bilayer} {Graphene}},\ }\href
  {https://doi.org/10.1103/PhysRevLett.121.037702} {\bibfield  {journal}
  {\bibinfo  {journal} {Phys. Rev. Lett.}\ }\textbf {\bibinfo {volume} {121}},\
  \bibinfo {pages} {037702} (\bibinfo {year} {2018})}\BibitemShut {NoStop}%
\bibitem [{\citenamefont {Tsim}\ \emph {et~al.}(2020)\citenamefont {Tsim},
  \citenamefont {Nam},\ and\ \citenamefont {Koshino}}]{tsim_perfect_2020}%
  \BibitemOpen
  \bibfield  {author} {\bibinfo {author} {\bibfnamefont {B.}~\bibnamefont
  {Tsim}}, \bibinfo {author} {\bibfnamefont {N.~N.~T.}\ \bibnamefont {Nam}},\
  and\ \bibinfo {author} {\bibfnamefont {M.}~\bibnamefont {Koshino}},\
  }\bibfield  {title} {\bibinfo {title} {Perfect one-dimensional chiral states
  in biased twisted bilayer graphene},\ }\href
  {https://doi.org/10.1103/PhysRevB.101.125409} {\bibfield  {journal} {\bibinfo
   {journal} {Phys. Rev. B}\ }\textbf {\bibinfo {volume} {101}},\ \bibinfo
  {pages} {125409} (\bibinfo {year} {2020})}\BibitemShut {NoStop}%
\bibitem [{\citenamefont {Efimkin}\ and\ \citenamefont
  {MacDonald}(2018)}]{efimkin_helical_2018}%
  \BibitemOpen
  \bibfield  {author} {\bibinfo {author} {\bibfnamefont {D.~K.}\ \bibnamefont
  {Efimkin}}\ and\ \bibinfo {author} {\bibfnamefont {A.~H.}\ \bibnamefont
  {MacDonald}},\ }\bibfield  {title} {\bibinfo {title} {Helical network model
  for twisted bilayer graphene},\ }\href
  {https://doi.org/10.1103/PhysRevB.98.035404} {\bibfield  {journal} {\bibinfo
  {journal} {Phys. Rev. B}\ }\textbf {\bibinfo {volume} {98}},\ \bibinfo
  {pages} {035404} (\bibinfo {year} {2018})}\BibitemShut {NoStop}%
\bibitem [{\citenamefont {Xu}\ \emph {et~al.}(2019)\citenamefont {Xu},
  \citenamefont {Berdyugin}, \citenamefont {Kumaravadivel}, \citenamefont
  {Guinea}, \citenamefont {Krishna~Kumar}, \citenamefont {Bandurin},
  \citenamefont {Morozov}, \citenamefont {Kuang}, \citenamefont {Tsim},
  \citenamefont {Liu}, \citenamefont {Edgar}, \citenamefont {Grigorieva},
  \citenamefont {Fal'ko}, \citenamefont {Kim},\ and\ \citenamefont
  {Geim}}]{xu_giant_2019}%
  \BibitemOpen
  \bibfield  {author} {\bibinfo {author} {\bibfnamefont {S.~G.}\ \bibnamefont
  {Xu}}, \bibinfo {author} {\bibfnamefont {A.~I.}\ \bibnamefont {Berdyugin}},
  \bibinfo {author} {\bibfnamefont {P.}~\bibnamefont {Kumaravadivel}}, \bibinfo
  {author} {\bibfnamefont {F.}~\bibnamefont {Guinea}}, \bibinfo {author}
  {\bibfnamefont {R.}~\bibnamefont {Krishna~Kumar}}, \bibinfo {author}
  {\bibfnamefont {D.~A.}\ \bibnamefont {Bandurin}}, \bibinfo {author}
  {\bibfnamefont {S.~V.}\ \bibnamefont {Morozov}}, \bibinfo {author}
  {\bibfnamefont {W.}~\bibnamefont {Kuang}}, \bibinfo {author} {\bibfnamefont
  {B.}~\bibnamefont {Tsim}}, \bibinfo {author} {\bibfnamefont {S.}~\bibnamefont
  {Liu}}, \bibinfo {author} {\bibfnamefont {J.~H.}\ \bibnamefont {Edgar}},
  \bibinfo {author} {\bibfnamefont {I.~V.}\ \bibnamefont {Grigorieva}},
  \bibinfo {author} {\bibfnamefont {V.~I.}\ \bibnamefont {Fal'ko}}, \bibinfo
  {author} {\bibfnamefont {M.}~\bibnamefont {Kim}},\ and\ \bibinfo {author}
  {\bibfnamefont {A.~K.}\ \bibnamefont {Geim}},\ }\bibfield  {title} {\bibinfo
  {title} {Giant oscillations in a triangular network of one-dimensional states
  in marginally twisted graphene},\ }\href
  {https://doi.org/10.1038/s41467-019-11971-7} {\bibfield  {journal} {\bibinfo
  {journal} {Nat. Commun.}\ }\textbf {\bibinfo {volume} {10}},\ \bibinfo
  {pages} {4008} (\bibinfo {year} {2019})}\BibitemShut {NoStop}%
\bibitem [{\citenamefont {Rickhaus}\ \emph {et~al.}(2018)\citenamefont
  {Rickhaus}, \citenamefont {Wallbank}, \citenamefont {Slizovskiy},
  \citenamefont {Pisoni}, \citenamefont {Overweg}, \citenamefont {Lee},
  \citenamefont {Eich}, \citenamefont {Liu}, \citenamefont {Watanabe},
  \citenamefont {Taniguchi}, \citenamefont {Ihn},\ and\ \citenamefont
  {Ensslin}}]{rickhaus_transport_2018}%
  \BibitemOpen
  \bibfield  {author} {\bibinfo {author} {\bibfnamefont {P.}~\bibnamefont
  {Rickhaus}}, \bibinfo {author} {\bibfnamefont {J.}~\bibnamefont {Wallbank}},
  \bibinfo {author} {\bibfnamefont {S.}~\bibnamefont {Slizovskiy}}, \bibinfo
  {author} {\bibfnamefont {R.}~\bibnamefont {Pisoni}}, \bibinfo {author}
  {\bibfnamefont {H.}~\bibnamefont {Overweg}}, \bibinfo {author} {\bibfnamefont
  {Y.}~\bibnamefont {Lee}}, \bibinfo {author} {\bibfnamefont {M.}~\bibnamefont
  {Eich}}, \bibinfo {author} {\bibfnamefont {M.-H.}\ \bibnamefont {Liu}},
  \bibinfo {author} {\bibfnamefont {K.}~\bibnamefont {Watanabe}}, \bibinfo
  {author} {\bibfnamefont {T.}~\bibnamefont {Taniguchi}}, \bibinfo {author}
  {\bibfnamefont {T.}~\bibnamefont {Ihn}},\ and\ \bibinfo {author}
  {\bibfnamefont {K.}~\bibnamefont {Ensslin}},\ }\bibfield  {title} {\bibinfo
  {title} {Transport {Through} a {Network} of {Topological} {Channels} in
  {Twisted} {Bilayer} {Graphene}},\ }\href
  {https://doi.org/10.1021/acs.nanolett.8b02387} {\bibfield  {journal}
  {\bibinfo  {journal} {Nano Lett.}\ }\textbf {\bibinfo {volume} {18}},\
  \bibinfo {pages} {6725} (\bibinfo {year} {2018})}\BibitemShut {NoStop}%
\bibitem [{\citenamefont {Aguilera}\ \emph {et~al.}(2020)\citenamefont
  {Aguilera}, \citenamefont {Jaeschke-Ubiergo}, \citenamefont {Vidal-Silva},
  \citenamefont {Torres},\ and\ \citenamefont {Nunez}}]{aguilera_topmagnon}%
  \BibitemOpen
  \bibfield  {author} {\bibinfo {author} {\bibfnamefont {E.}~\bibnamefont
  {Aguilera}}, \bibinfo {author} {\bibfnamefont {R.}~\bibnamefont
  {Jaeschke-Ubiergo}}, \bibinfo {author} {\bibfnamefont {N.}~\bibnamefont
  {Vidal-Silva}}, \bibinfo {author} {\bibfnamefont {L.~E. F.~F.}\ \bibnamefont
  {Torres}},\ and\ \bibinfo {author} {\bibfnamefont {A.~S.}\ \bibnamefont
  {Nunez}},\ }\bibfield  {title} {\bibinfo {title} {Topological magnonics in
  the two-dimensional van der {Waals} magnet {CrI}$_3$},\ }\href
  {https://doi.org/10.1103/PhysRevB.102.024409} {\bibfield  {journal} {\bibinfo
   {journal} {Phys. Rev. B}\ }\textbf {\bibinfo {volume} {102}},\ \bibinfo
  {pages} {024409} (\bibinfo {year} {2020})}\BibitemShut {NoStop}%
\bibitem [{\citenamefont {Costa}\ \emph {et~al.}(2020)\citenamefont {Costa},
  \citenamefont {Santos}, \citenamefont {Peres},\ and\ \citenamefont
  {Fernández-Rossier}}]{costa_topological_2020}%
  \BibitemOpen
  \bibfield  {author} {\bibinfo {author} {\bibfnamefont {A.~T.}\ \bibnamefont
  {Costa}}, \bibinfo {author} {\bibfnamefont {D.~L.~R.}\ \bibnamefont
  {Santos}}, \bibinfo {author} {\bibfnamefont {N.~M.~R.}\ \bibnamefont
  {Peres}},\ and\ \bibinfo {author} {\bibfnamefont {J.}~\bibnamefont
  {Fernández-Rossier}},\ }\bibfield  {title} {\bibinfo {title} {Topological
  magnons in {CrI}$_3$ monolayers: an itinerant fermion description},\ }\href
  {https://doi.org/10.1088/2053-1583/aba88f} {\bibfield  {journal} {\bibinfo
  {journal} {2D Mater.}\ }\textbf {\bibinfo {volume} {7}},\ \bibinfo {pages}
  {045031} (\bibinfo {year} {2020})}\BibitemShut {NoStop}%
\bibitem [{\citenamefont {Kresse}\ and\ \citenamefont
  {Furthm\"uller}(1996)}]{vasp1}%
  \BibitemOpen
  \bibfield  {author} {\bibinfo {author} {\bibfnamefont {G.}~\bibnamefont
  {Kresse}}\ and\ \bibinfo {author} {\bibfnamefont {J.}~\bibnamefont
  {Furthm\"uller}},\ }\bibfield  {title} {\bibinfo {title} {Efficient iterative
  schemes for ab initio total-energy calculations using a plane-wave basis
  set},\ }\href {https://doi.org/10.1103/PhysRevB.54.11169} {\bibfield
  {journal} {\bibinfo  {journal} {Phys. Rev. B}\ }\textbf {\bibinfo {volume}
  {54}},\ \bibinfo {pages} {11169} (\bibinfo {year} {1996})}\BibitemShut
  {NoStop}%
\bibitem [{\citenamefont {Kresse}\ and\ \citenamefont {Joubert}(1999)}]{vasp2}%
  \BibitemOpen
  \bibfield  {author} {\bibinfo {author} {\bibfnamefont {G.}~\bibnamefont
  {Kresse}}\ and\ \bibinfo {author} {\bibfnamefont {D.}~\bibnamefont
  {Joubert}},\ }\bibfield  {title} {\bibinfo {title} {From ultrasoft
  pseudopotentials to the projector augmented-wave method},\ }\href
  {https://doi.org/10.1103/PhysRevB.59.1758} {\bibfield  {journal} {\bibinfo
  {journal} {Phys. Rev. B}\ }\textbf {\bibinfo {volume} {59}},\ \bibinfo
  {pages} {1758} (\bibinfo {year} {1999})}\BibitemShut {NoStop}%
\bibitem [{\citenamefont {Bl\"ochl}(1994)}]{blochl_projector_1994}%
  \BibitemOpen
  \bibfield  {author} {\bibinfo {author} {\bibfnamefont {P.~E.}\ \bibnamefont
  {Bl\"ochl}},\ }\bibfield  {title} {\bibinfo {title} {Projector augmented-wave
  method},\ }\href {https://doi.org/10.1103/PhysRevB.50.17953} {\bibfield
  {journal} {\bibinfo  {journal} {Phys. Rev. B}\ }\textbf {\bibinfo {volume}
  {50}},\ \bibinfo {pages} {17953} (\bibinfo {year} {1994})}\BibitemShut
  {NoStop}%
\bibitem [{\citenamefont {Csonka}\ \emph {et~al.}(2009)\citenamefont {Csonka},
  \citenamefont {Perdew}, \citenamefont {Ruzsinszky}, \citenamefont
  {Philipsen}, \citenamefont {Leb\`egue}, \citenamefont {Paier}, \citenamefont
  {Vydrov},\ and\ \citenamefont {\'Angy\'an}}]{PBEsol}%
  \BibitemOpen
  \bibfield  {author} {\bibinfo {author} {\bibfnamefont {G.~I.}\ \bibnamefont
  {Csonka}}, \bibinfo {author} {\bibfnamefont {J.~P.}\ \bibnamefont {Perdew}},
  \bibinfo {author} {\bibfnamefont {A.}~\bibnamefont {Ruzsinszky}}, \bibinfo
  {author} {\bibfnamefont {P.~H.~T.}\ \bibnamefont {Philipsen}}, \bibinfo
  {author} {\bibfnamefont {S.}~\bibnamefont {Leb\`egue}}, \bibinfo {author}
  {\bibfnamefont {J.}~\bibnamefont {Paier}}, \bibinfo {author} {\bibfnamefont
  {O.~A.}\ \bibnamefont {Vydrov}},\ and\ \bibinfo {author} {\bibfnamefont
  {J.~G.}\ \bibnamefont {\'Angy\'an}},\ }\bibfield  {title} {\bibinfo {title}
  {Assessing the performance of recent density functionals for bulk solids},\
  }\href {https://doi.org/10.1103/PhysRevB.79.155107} {\bibfield  {journal}
  {\bibinfo  {journal} {Phys. Rev. B}\ }\textbf {\bibinfo {volume} {79}},\
  \bibinfo {pages} {155107} (\bibinfo {year} {2009})}\BibitemShut {NoStop}%
\bibitem [{\citenamefont {Bezanson}\ \emph {et~al.}(2017)\citenamefont
  {Bezanson}, \citenamefont {Edelman}, \citenamefont {Karpinski},\ and\
  \citenamefont {Shah}}]{bezanson2017julia}%
  \BibitemOpen
  \bibfield  {author} {\bibinfo {author} {\bibfnamefont {J.}~\bibnamefont
  {Bezanson}}, \bibinfo {author} {\bibfnamefont {A.}~\bibnamefont {Edelman}},
  \bibinfo {author} {\bibfnamefont {S.}~\bibnamefont {Karpinski}},\ and\
  \bibinfo {author} {\bibfnamefont {V.~B.}\ \bibnamefont {Shah}},\ }\bibfield
  {title} {\bibinfo {title} {Julia: A fresh approach to numerical computing},\
  }\href@noop {} {\bibfield  {journal} {\bibinfo  {journal} {SIAM Rev.}\
  }\textbf {\bibinfo {volume} {59}},\ \bibinfo {pages} {65} (\bibinfo {year}
  {2017})}\BibitemShut {NoStop}%
\bibitem [{\citenamefont {Rackauckas}\ and\ \citenamefont
  {Nie}(2017)}]{rackauckas2017differentialequations}%
  \BibitemOpen
  \bibfield  {author} {\bibinfo {author} {\bibfnamefont {C.}~\bibnamefont
  {Rackauckas}}\ and\ \bibinfo {author} {\bibfnamefont {Q.}~\bibnamefont
  {Nie}},\ }\bibfield  {title} {\bibinfo {title} {Differentialequations.jl -- a
  performant and feature-rich ecosystem for solving differential equations in
  julia},\ }\href@noop {} {\bibfield  {journal} {\bibinfo  {journal} {J. Open
  Res. Softw.}\ }\textbf {\bibinfo {volume} {5}} (\bibinfo {year}
  {2017})}\BibitemShut {NoStop}%
\bibitem [{\citenamefont {Gargiulo}\ and\ \citenamefont
  {Yazyev}(2017)}]{gargiulo_structural_2017}%
  \BibitemOpen
  \bibfield  {author} {\bibinfo {author} {\bibfnamefont {F.}~\bibnamefont
  {Gargiulo}}\ and\ \bibinfo {author} {\bibfnamefont {O.~V.}\ \bibnamefont
  {Yazyev}},\ }\bibfield  {title} {\bibinfo {title} {Structural and electronic
  transformation in low-angle twisted bilayer graphene},\ }\href
  {https://doi.org/10.1088/2053-1583/aa9640} {\bibfield  {journal} {\bibinfo
  {journal} {2D Mater.}\ }\textbf {\bibinfo {volume} {5}},\ \bibinfo {pages}
  {015019} (\bibinfo {year} {2017})}\BibitemShut {NoStop}%
\bibitem [{\citenamefont {Grimme}\ \emph {et~al.}(2010)\citenamefont {Grimme},
  \citenamefont {Antony}, \citenamefont {Ehrlich},\ and\ \citenamefont
  {Krieg}}]{DFT_D3}%
  \BibitemOpen
  \bibfield  {author} {\bibinfo {author} {\bibfnamefont {S.}~\bibnamefont
  {Grimme}}, \bibinfo {author} {\bibfnamefont {J.}~\bibnamefont {Antony}},
  \bibinfo {author} {\bibfnamefont {S.}~\bibnamefont {Ehrlich}},\ and\ \bibinfo
  {author} {\bibfnamefont {H.}~\bibnamefont {Krieg}},\ }\bibfield  {title}
  {\bibinfo {title} {A consistent and accurate ab initio parametrization of
  density functional dispersion correction ({DFT-D}) for the 94 elements
  {H-Pu}},\ }\href@noop {} {\bibfield  {journal} {\bibinfo  {journal} {J. Chem.
  Phys.}\ }\textbf {\bibinfo {volume} {132}},\ \bibinfo {pages} {154104}
  (\bibinfo {year} {2010})}\BibitemShut {NoStop}%
\bibitem [{\citenamefont {Zhang}\ and\ \citenamefont
  {Zhang}(2012)}]{zhang_spin_2012}%
  \BibitemOpen
  \bibfield  {author} {\bibinfo {author} {\bibfnamefont {S.~S.-L.}\
  \bibnamefont {Zhang}}\ and\ \bibinfo {author} {\bibfnamefont
  {S.}~\bibnamefont {Zhang}},\ }\bibfield  {title} {\bibinfo {title} {Spin
  convertance at magnetic interfaces},\ }\href
  {https://doi.org/10.1103/PhysRevB.86.214424} {\bibfield  {journal} {\bibinfo
  {journal} {Phys. Rev. B}\ }\textbf {\bibinfo {volume} {86}},\ \bibinfo
  {pages} {214424} (\bibinfo {year} {2012})}\BibitemShut {NoStop}%
\end{thebibliography}%


\begin{thebibliography}{10}%
\makeatletter
\providecommand \@ifxundefined [1]{%
 \@ifx{#1\undefined}
}%
\providecommand \@ifnum [1]{%
 \ifnum #1\expandafter \@firstoftwo
 \else \expandafter \@secondoftwo
 \fi
}%
\providecommand \@ifx [1]{%
 \ifx #1\expandafter \@firstoftwo
 \else \expandafter \@secondoftwo
 \fi
}%
\providecommand \natexlab [1]{#1}%
\providecommand \enquote  [1]{``#1''}%
\providecommand \bibnamefont  [1]{#1}%
\providecommand \bibfnamefont [1]{#1}%
\providecommand \citenamefont [1]{#1}%
\providecommand \href@noop [0]{\@secondoftwo}%
\providecommand \href [0]{\begingroup \@sanitize@url \@href}%
\providecommand \@href[1]{\@@startlink{#1}\@@href}%
\providecommand \@@href[1]{\endgroup#1\@@endlink}%
\providecommand \@sanitize@url [0]{\catcode `\\12\catcode `\$12\catcode
  `\&12\catcode `\#12\catcode `\^12\catcode `\_12\catcode `\%12\relax}%
\providecommand \@@startlink[1]{}%
\providecommand \@@endlink[0]{}%
\providecommand \url  [0]{\begingroup\@sanitize@url \@url }%
\providecommand \@url [1]{\endgroup\@href {#1}{\urlprefix }}%
\providecommand \urlprefix  [0]{URL }%
\providecommand \Eprint [0]{\href }%
\providecommand \doibase [0]{https://doi.org/}%
\providecommand \selectlanguage [0]{\@gobble}%
\providecommand \bibinfo  [0]{\@secondoftwo}%
\providecommand \bibfield  [0]{\@secondoftwo}%
\providecommand \translation [1]{[#1]}%
\providecommand \BibitemOpen [0]{}%
\providecommand \bibitemStop [0]{}%
\providecommand \bibitemNoStop [0]{.\EOS\space}%
\providecommand \EOS [0]{\spacefactor3000\relax}%
\providecommand \BibitemShut  [1]{\csname bibitem#1\endcsname}%
\let\auto@bib@innerbib\@empty
\bibitem [{\citenamefont {Kresse}\ and\ \citenamefont
  {Furthm\"uller}(1996)}]{vasp1}%
  \BibitemOpen
  \bibfield  {author} {\bibinfo {author} {\bibfnamefont {G.}~\bibnamefont
  {Kresse}}\ and\ \bibinfo {author} {\bibfnamefont {J.}~\bibnamefont
  {Furthm\"uller}},\ }\bibfield  {title} {\bibinfo {title} {Efficient iterative
  schemes for ab initio total-energy calculations using a plane-wave basis
  set},\ }\href {https://doi.org/10.1103/PhysRevB.54.11169} {\bibfield
  {journal} {\bibinfo  {journal} {Phys. Rev. B}\ }\textbf {\bibinfo {volume}
  {54}},\ \bibinfo {pages} {11169} (\bibinfo {year} {1996})}\BibitemShut
  {NoStop}%
\bibitem [{\citenamefont {Kresse}\ and\ \citenamefont {Joubert}(1999)}]{vasp2}%
  \BibitemOpen
  \bibfield  {author} {\bibinfo {author} {\bibfnamefont {G.}~\bibnamefont
  {Kresse}}\ and\ \bibinfo {author} {\bibfnamefont {D.}~\bibnamefont
  {Joubert}},\ }\bibfield  {title} {\bibinfo {title} {From ultrasoft
  pseudopotentials to the projector augmented-wave method},\ }\href
  {https://doi.org/10.1103/PhysRevB.59.1758} {\bibfield  {journal} {\bibinfo
  {journal} {Phys. Rev. B}\ }\textbf {\bibinfo {volume} {59}},\ \bibinfo
  {pages} {1758} (\bibinfo {year} {1999})}\BibitemShut {NoStop}%
\bibitem [{\citenamefont {Bl\"ochl}(1994)}]{blochl_projector_1994}%
  \BibitemOpen
  \bibfield  {author} {\bibinfo {author} {\bibfnamefont {P.~E.}\ \bibnamefont
  {Bl\"ochl}},\ }\bibfield  {title} {\bibinfo {title} {Projector augmented-wave
  method},\ }\href {https://doi.org/10.1103/PhysRevB.50.17953} {\bibfield
  {journal} {\bibinfo  {journal} {Phys. Rev. B}\ }\textbf {\bibinfo {volume}
  {50}},\ \bibinfo {pages} {17953} (\bibinfo {year} {1994})}\BibitemShut
  {NoStop}%
\bibitem [{\citenamefont {Csonka}\ \emph {et~al.}(2009)\citenamefont {Csonka},
  \citenamefont {Perdew}, \citenamefont {Ruzsinszky}, \citenamefont
  {Philipsen}, \citenamefont {Leb\`egue}, \citenamefont {Paier}, \citenamefont
  {Vydrov},\ and\ \citenamefont {\'Angy\'an}}]{PBEsol}%
  \BibitemOpen
  \bibfield  {author} {\bibinfo {author} {\bibfnamefont {G.~I.}\ \bibnamefont
  {Csonka}}, \bibinfo {author} {\bibfnamefont {J.~P.}\ \bibnamefont {Perdew}},
  \bibinfo {author} {\bibfnamefont {A.}~\bibnamefont {Ruzsinszky}}, \bibinfo
  {author} {\bibfnamefont {P.~H.~T.}\ \bibnamefont {Philipsen}}, \bibinfo
  {author} {\bibfnamefont {S.}~\bibnamefont {Leb\`egue}}, \bibinfo {author}
  {\bibfnamefont {J.}~\bibnamefont {Paier}}, \bibinfo {author} {\bibfnamefont
  {O.~A.}\ \bibnamefont {Vydrov}},\ and\ \bibinfo {author} {\bibfnamefont
  {J.~G.}\ \bibnamefont {\'Angy\'an}},\ }\bibfield  {title} {\bibinfo {title}
  {Assessing the performance of recent density functionals for bulk solids},\
  }\href {https://doi.org/10.1103/PhysRevB.79.155107} {\bibfield  {journal}
  {\bibinfo  {journal} {Phys. Rev. B}\ }\textbf {\bibinfo {volume} {79}},\
  \bibinfo {pages} {155107} (\bibinfo {year} {2009})}\BibitemShut {NoStop}%
\bibitem [{\citenamefont {Grimme}\ \emph {et~al.}(2010)\citenamefont {Grimme},
  \citenamefont {Antony}, \citenamefont {Ehrlich},\ and\ \citenamefont
  {Krieg}}]{DFT_D3}%
  \BibitemOpen
  \bibfield  {author} {\bibinfo {author} {\bibfnamefont {S.}~\bibnamefont
  {Grimme}}, \bibinfo {author} {\bibfnamefont {J.}~\bibnamefont {Antony}},
  \bibinfo {author} {\bibfnamefont {S.}~\bibnamefont {Ehrlich}},\ and\ \bibinfo
  {author} {\bibfnamefont {H.}~\bibnamefont {Krieg}},\ }\bibfield  {title}
  {\bibinfo {title} {A consistent and accurate ab initio parametrization of
  density functional dispersion correction ({DFT-D}) for the 94 elements
  {H-Pu}},\ }\href@noop {} {\bibfield  {journal} {\bibinfo  {journal} {J. Chem.
  Phys.}\ }\textbf {\bibinfo {volume} {132}},\ \bibinfo {pages} {154104}
  (\bibinfo {year} {2010})}\BibitemShut {NoStop}%
\bibitem [{\citenamefont {Carr}\ \emph {et~al.}(2018)\citenamefont {Carr},
  \citenamefont {Massatt}, \citenamefont {Torrisi}, \citenamefont {Cazeaux},
  \citenamefont {Luskin},\ and\ \citenamefont
  {Kaxiras}}]{carr_relaxation_2018}%
  \BibitemOpen
  \bibfield  {author} {\bibinfo {author} {\bibfnamefont {S.}~\bibnamefont
  {Carr}}, \bibinfo {author} {\bibfnamefont {D.}~\bibnamefont {Massatt}},
  \bibinfo {author} {\bibfnamefont {S.~B.}\ \bibnamefont {Torrisi}}, \bibinfo
  {author} {\bibfnamefont {P.}~\bibnamefont {Cazeaux}}, \bibinfo {author}
  {\bibfnamefont {M.}~\bibnamefont {Luskin}},\ and\ \bibinfo {author}
  {\bibfnamefont {E.}~\bibnamefont {Kaxiras}},\ }\bibfield  {title} {\bibinfo
  {title} {Relaxation and domain formation in incommensurate two-dimensional
  heterostructures},\ }\href {https://doi.org/10.1103/PhysRevB.98.224102}
  {\bibfield  {journal} {\bibinfo  {journal} {Phys. Rev. B}\ }\textbf {\bibinfo
  {volume} {98}},\ \bibinfo {pages} {224102} (\bibinfo {year}
  {2018})}\BibitemShut {NoStop}%
\bibitem [{\citenamefont {Bezanson}\ \emph {et~al.}(2017)\citenamefont
  {Bezanson}, \citenamefont {Edelman}, \citenamefont {Karpinski},\ and\
  \citenamefont {Shah}}]{bezanson2017julia}%
  \BibitemOpen
  \bibfield  {author} {\bibinfo {author} {\bibfnamefont {J.}~\bibnamefont
  {Bezanson}}, \bibinfo {author} {\bibfnamefont {A.}~\bibnamefont {Edelman}},
  \bibinfo {author} {\bibfnamefont {S.}~\bibnamefont {Karpinski}},\ and\
  \bibinfo {author} {\bibfnamefont {V.~B.}\ \bibnamefont {Shah}},\ }\bibfield
  {title} {\bibinfo {title} {Julia: A fresh approach to numerical computing},\
  }\href@noop {} {\bibfield  {journal} {\bibinfo  {journal} {SIAM Rev.}\
  }\textbf {\bibinfo {volume} {59}},\ \bibinfo {pages} {65} (\bibinfo {year}
  {2017})}\BibitemShut {NoStop}%
\bibitem [{\citenamefont {Rackauckas}\ and\ \citenamefont
  {Nie}(2017)}]{rackauckas2017differentialequations}%
  \BibitemOpen
  \bibfield  {author} {\bibinfo {author} {\bibfnamefont {C.}~\bibnamefont
  {Rackauckas}}\ and\ \bibinfo {author} {\bibfnamefont {Q.}~\bibnamefont
  {Nie}},\ }\bibfield  {title} {\bibinfo {title} {Differentialequations.jl -- a
  performant and feature-rich ecosystem for solving differential equations in
  julia},\ }\href@noop {} {\bibfield  {journal} {\bibinfo  {journal} {J. Open
  Res. Softw.}\ }\textbf {\bibinfo {volume} {5}} (\bibinfo {year}
  {2017})}\BibitemShut {NoStop}%
\bibitem [{\citenamefont {Gargiulo}\ and\ \citenamefont
  {Yazyev}(2017)}]{gargiulo_structural_2017}%
  \BibitemOpen
  \bibfield  {author} {\bibinfo {author} {\bibfnamefont {F.}~\bibnamefont
  {Gargiulo}}\ and\ \bibinfo {author} {\bibfnamefont {O.~V.}\ \bibnamefont
  {Yazyev}},\ }\bibfield  {title} {\bibinfo {title} {Structural and electronic
  transformation in low-angle twisted bilayer graphene},\ }\href
  {https://doi.org/10.1088/2053-1583/aa9640} {\bibfield  {journal} {\bibinfo
  {journal} {2D Mater.}\ }\textbf {\bibinfo {volume} {5}},\ \bibinfo {pages}
  {015019} (\bibinfo {year} {2017})}\BibitemShut {NoStop}%
\bibitem [{\citenamefont {Zhang}\ and\ \citenamefont
  {Zhang}(2012)}]{zhang_spin_2012}%
  \BibitemOpen
  \bibfield  {author} {\bibinfo {author} {\bibfnamefont {S.~S.-L.}\
  \bibnamefont {Zhang}}\ and\ \bibinfo {author} {\bibfnamefont
  {S.}~\bibnamefont {Zhang}},\ }\bibfield  {title} {\bibinfo {title} {Spin
  convertance at magnetic interfaces},\ }\href
  {https://doi.org/10.1103/PhysRevB.86.214424} {\bibfield  {journal} {\bibinfo
  {journal} {Phys. Rev. B}\ }\textbf {\bibinfo {volume} {86}},\ \bibinfo
  {pages} {214424} (\bibinfo {year} {2012})}\BibitemShut {NoStop}%
\end{thebibliography}%

\end{document}


\title{Supplemental Material for `Stacking Domain Wall Magnons in Twisted van der Waals Magnets'}

\author{Chong \surname{Wang}}
\affiliation{Department of Physics, Carnegie Mellon University, Pittsburgh, Pennsylvania 15213, USA}

\author{Yuan \surname{Gao}}
\affiliation{Department of Physics, Carnegie Mellon University, Pittsburgh, Pennsylvania 15213, USA}
\affiliation{International Center for Quantum Design of Functional Materials (ICQD), Hefei National Laboratory for Physical Sciences at the
Microscale, University of Science and Technology of China, Hefei, Anhui 230026, China}

\author{Hongyan \surname{Lv}}
\affiliation{Department of Physics, Carnegie Mellon University, Pittsburgh, Pennsylvania 15213, USA}
\affiliation{Key Laboratory of Materials Physics, Institute of Solid State Physics, Chinese Academy of Sciences, 
Hefei, 230031, China}

\author{Xiaodong \surname{Xu}}
\affiliation{Department of Physics, University of Washington, Seattle, Washington 98195, USA}
\affiliation{Department of Materials Science and Engineering, University of Washington, Seattle, Washington 98195, USA}

\author{Di \surname{Xiao}}
\affiliation{Department of Physics, Carnegie Mellon University, Pittsburgh, Pennsylvania 15213, USA}

\maketitle

\section{DFT methods}

The calculations of interlayer potential energy, interlayer and intralayer exchange coupling, monolayer magnetic anisotropy, bulk modulus, shear modulus are carried out within the framework of density functional theory (DFT).  DFT calculations are performed with the Vienna Ab initio Simulation Package~\cite{vasp1,vasp2} using projector augmented wave (PAW) potential~\cite{blochl_projector_1994}. The electron-electron interaction is described by the PBEsol exchange-correlation functional~\cite{PBEsol}. The plane wave basis cutoff is 500 eV and sampling mesh in reciprocal space is $8\times 8 \times 1$. The atomic structures are relaxed until forces on every atom are smaller than 0.005 eV/\AA~and a vacuum layer larger than 25 \AA~is inserted between the periodic images. A Hubbard $U=3$ eV was added to the $d$ orbitals of Cr to capture the local Coulomb interaction. The van der Waals functional used here is zero damping DFT-D3 method \cite{DFT_D3} with small modifications: we here only take into account the interactions between the atoms from two different layers. This modification has the advantage that it gives rise to the same lattice constant for monolayers and bilayers in the relaxation calculations.

The bulk modulus and shear modulus is fitted to the DFT total energy on a $11\times 11$ mesh, where the lattice are stretched from -5\% to 5\% with a step of 1\% in the in-plane directions.

\section{Lattice relaxation}

\begin{figure*}
\centering
\includegraphics[width=1\textwidth]{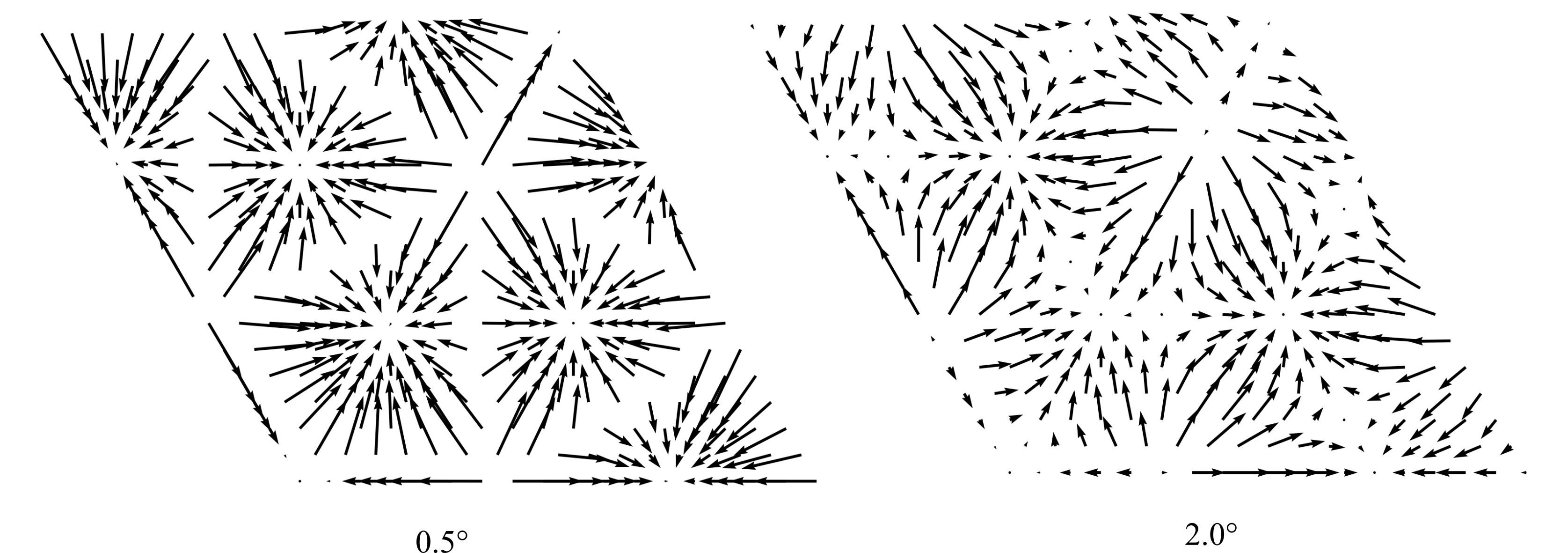}
\caption{Lattice relaxation of twisted bilayer CrI$_3$ with twist angle 0.5$^\circ$ and 2.0$^\circ$. The arrows represent the displacement $\bm{u}(\bm{d})$.\label{fig:displacemnts}}
\end{figure*}

For convenience, in this section, we are choosing a different coordinate from the main text: $z$ is the out-of-plane direction and $x$, $y$ are in-plane directions.

Continuous theory assumes the energy of the Moir\'e structure can be expressed as a real space integral of local energy density. To express the energy density in a way that is suitable for numerical optimization, we follow the practice of Ref.~\onlinecite{carr_relaxation_2018}. For small twist angle $\theta$, the unrelaxed twisted CrI$_3$ at point $\bm{r}$ resembles a bilayer CrI$_3$ where the top layer is \textit{translated} with respect to the bottom layer by $\bm{d}(\bm{r})=\theta\hat{\bm{z}}\times\bm{r}$. Here, the vector $\bm{d}$ will be referred to as the stacking vector before relaxation. Since lattice translations do not change the local configuration, $\bm{d}$ is only defined modulo lattice translations and can be made to take value from the unit cell of the Bravais lattice. Therefore, the unit cell will be dubbed as the stacking space. In the regime of continuous theory, relaxation is locally viewed as translation and is described by two displacement field  $\bm{u}^{t}(\bm{r})$ (top layer) and $\bm{u}^{b}(\bm{r})$ (bottom layer), changing the stacking vector to $\bm{b}(\bm{r})=\bm{d}(\bm{r})+\bm{u}^t(\bm{r})-\bm{u}^b(\bm{r})$. We further assume $\bm{u}^{t(b)}(\bm{r})$ only depends on $\bm{r}$ through $\bm{d}$ such that the relaxation is completely determined by $\bm{u}^{t(b)}$ as a function of $\bm{d}$. For incommensurate Moir\'e structure, the distribution of $\bm{d}(\bm{r})$ is uniform in the stacking space. Therefore, the interlayer potential energy per unit cell can be obtained by an integration in the stacking space.
\begin{equation}
E_\mathrm{pot} = \frac{1}{|P|}\int_P V[\bm{d}+\bm{u}^t(\bm{d})-\bm{u}^b(\bm{d})]\mathrm{d}\bm{d},\label{pot}
\end{equation}
where $P$ is the unit cell with $|P|$ being its area, and $V[\bm{b}]$ is the interlayer potential energy per unit cell for the bilayer CrI$_3$ with stacking order $\bm{b}$.

The intralayer energy is modeled by elastic theory. Since the point group of monolayer CrI$_3$ is $D_{3d}$, the elastic stiffness tensor contains only two independent parameters: bulk modulus $B$ and shear modulus $G$. Although the symmetry of the monolayer would be lowered by magnetism, we will assume such effects can be neglected in the elastic energy. Similar to the interlayer potential energy, the intralayer elastic energy per unit cell can also be written as an integration in the stacking space
\begin{widetext}
\begin{equation}
E_\mathrm{elastic}^l=\frac{\theta^2}{2|P|}\int_P \biggl\{K\left(\frac{\partial u_x^l}{\partial d_y}-\frac{\partial u_y^l}{\partial d_x}\right)^2 + G\left[\left(\frac{\partial u_x^l}{\partial d_y}+\frac{\partial u_y^l}{\partial d_x}\right)^2+\left(\frac{\partial u_x^l}{\partial d_x}-\frac{\partial u_y^l}{\partial d_y}\right)^2\right]\biggr\}\mathrm{d}\bm{d}, \label{elastic}
\end{equation}
\end{widetext}
where $l\in \{t,b\}$ is the layer index.

Minimization of  $E_\mathrm{relax} = E_\mathrm{pot}+E_\mathrm{elastic}^t+E_\mathrm{elastic}^b$ can be done by solving the Euler-Lagrangian equation $\delta E_\mathrm{relax}/\delta \bm{u} = \bm{0}$. Solution of Euler-Lagrangian equation is in turn obtained numerically by discretizing $\bm{u}$ in the unit cell P in a $18 \times 18$ mesh and then solving the following differential equation 
\begin{equation}
    \frac{\dd\bm{u}}{\dd t} = - \frac{\delta E_\mathrm{relax}}{\delta \bm{u}}.
\end{equation}
The equation above will converge to a stable state, which is a solution to the Euler-Lagrangian equation. \textsc{julia} programming language~\cite{bezanson2017julia} and \textsc{DifferentialEquations.jl} package~\cite{rackauckas2017differentialequations} have been utilized to solve the differential equation.

The results of the minimization for 0.5$^\circ$ and 2.0$^\circ$ is plotted in Fig.~\ref{fig:displacemnts}. As expected, stackings with higher energies will generally flow to stackings with lower energies.  This flow is more evident for small angles due to their weaker elastic energy cost [Eq.~\eqref{elastic} is proportional to $\theta^2$].

It is worth noting that besides the two stable stackings with roughly the same energy, there is an additional local minimum -- the AA stacking -- labelled by green dots in Fig. 1(a) in the main text. The potential energy landscape is highly anisotropic around the AA stacking. In the intermediate twist angles, there will be an additional AA-like domain in the relaxed structure.  AA stacking bilayer CrI$_3$ does not exist in nature, but as a local energy minimum, we envision its presence in twisted bilayer CrI$_3$ will enrich its optical and transport properties.

\begin{figure}[b]
\centering
\includegraphics[width=1.0\columnwidth]{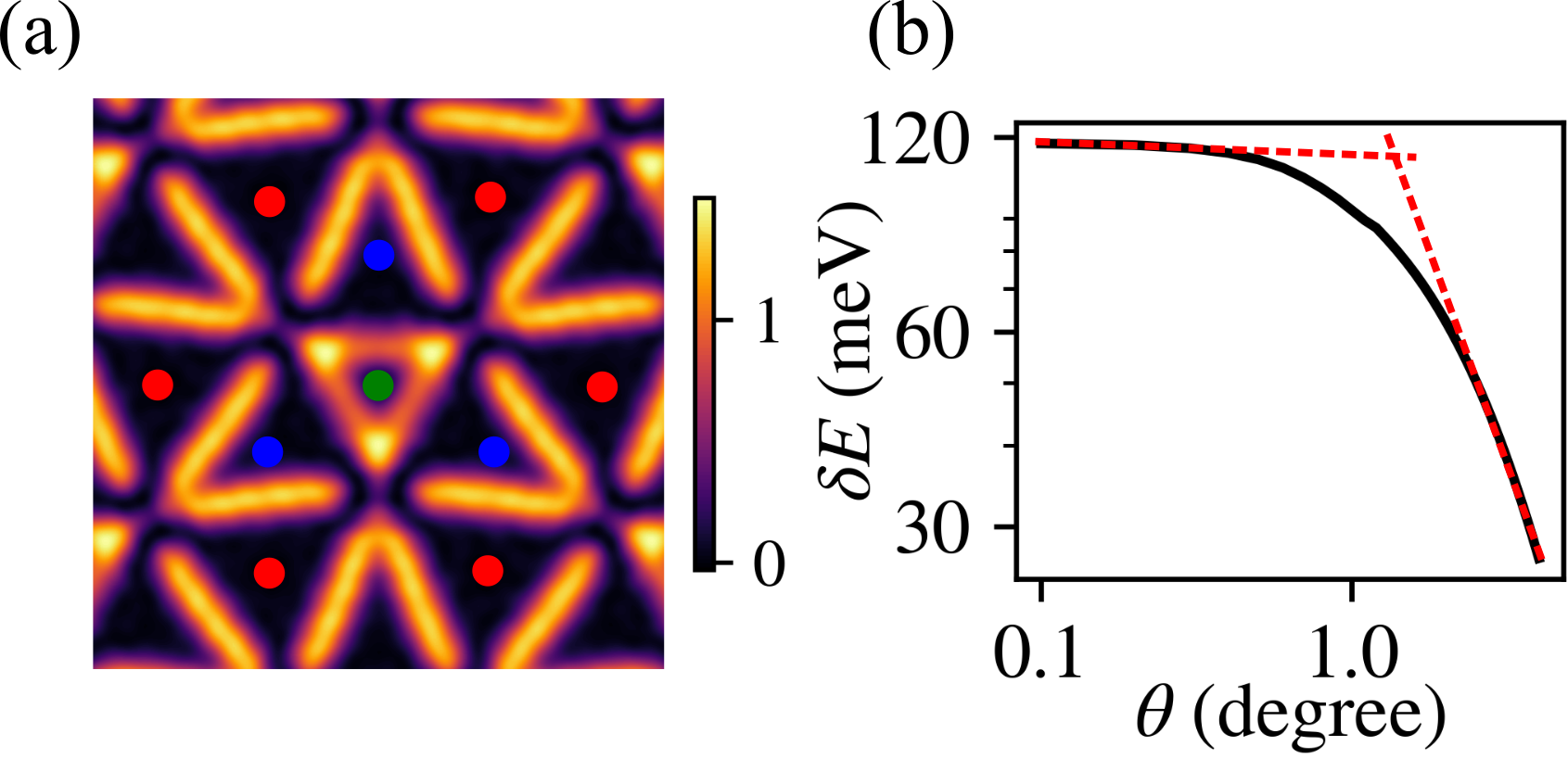}
\caption{(a) Shear strain in the real space for twist angle 2.0$^\circ$. Unit: $10^{-3}$. The color dots follow the convention in the main text. (b) The log scale plot of energy reduction with respect to the twist angle. The red dashed lines are used to estimate the angle at which the domains are formed.\label{fig:domain}}
\end{figure}

In twisted bilayers, structural domain walls  generally bear shear strain density. To visualize the structural domain in the real space, we plot in Fig.~\ref{fig:domain}(a) the real space map of shear strain density. All stacking domains are labelled with color dots. AA-like domain can be clearly observed in the picture.

With decreasing twist angle, the energy reduction per unit cell due to relaxation increases. This reduction of energy has an upper bound: in the limit of zero twist angle, the reduction of energy per unit cell approaches the interlayer potential energy averaged with respect to different stackings $\int_P \mathrm{d}\boldsymbol{b} V(\boldsymbol{b})/|P|$ minus the energy per unit cell for rhombohedral and monoclinic stackings. This limiting behavior is illustrated in Fig. \ref{fig:domain}(b). Such a behavior makes it possible for Ref.~\onlinecite{gargiulo_structural_2017} to define a critical angle when domains are formed. This angle in the current context is estimated to be 1.3$^\circ$ degree. When domains are fully developed, the structure of the Moire lattice are more flexible, in the sense that the energy cost of local deformation is small and the domain walls receive some freedom to move about. Correspondingly, we observe that the relaxed $\boldsymbol{u}$ begin to depend sensitively on the initial configuration at 1.1$^\circ$ in the minimization of the energy density.

\section{Magnetic properties of MM domain wall}

\begin{figure}
\centering
\includegraphics[width=1.0\columnwidth]{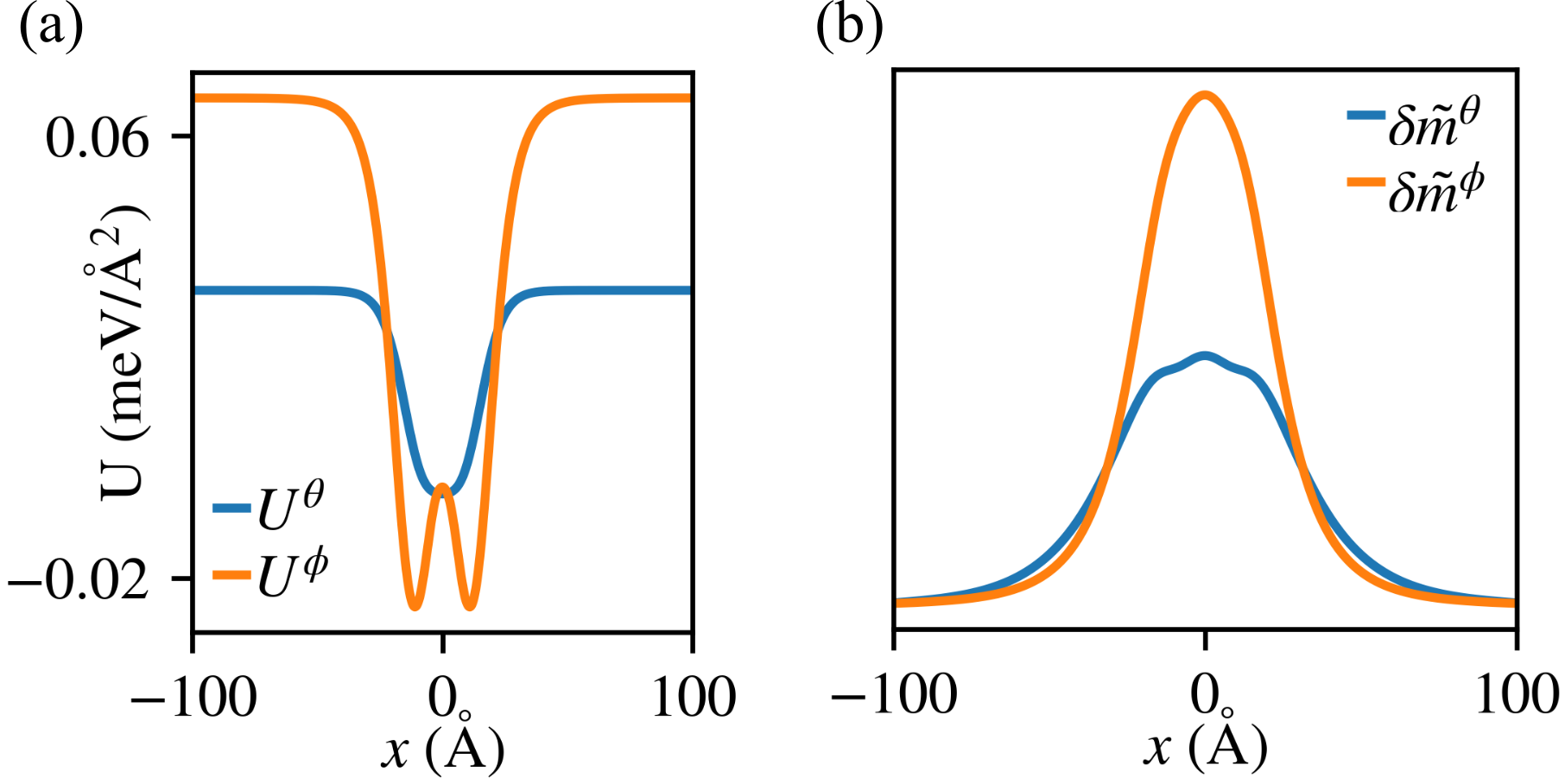}
\caption{{(a) The trapping potentials $U^\theta$ and $U^\phi$ for $\delta \bm{\tm}_+$.} (b) Bounded magnon at $k_z=0$ for $\delta \bm{\tm}_-$.\label{fig:MMmagnon_supp}}
\end{figure}

We choose the coordinates such that the stacking domain wall lies in the $z$
direction at $x \sim 0$. $y$ is the out-of-plane direction. We assume the
energy is described by the following energy functional
\begin{eqnarray*}
  E & = & \int \sum_{\alpha, \beta, l} \frac{A}{2} (\partial_{\beta}
  m_l^{\alpha})^2 - \sum_l \frac{K}{2}  (m_l^y)^2 - \sum_{\alpha}
  Jm_1^{\alpha} m_2^{\alpha} \mathd x,
\end{eqnarray*}
where $A$ and $K$ are independent of $x$; $\tmmathbf{m}_l$ and $J$ are
dependent of $x$; $\tmmathbf{m}$ has unit length. $A$ is positive to suppress
magnetization fluctuation; $K$ is positive such that $\tmmathbf{m}$ favors
out-of-plane direction. By variation, the effective magnetic field reads
($\bar{1} = 2$ and $\bar{2} = 1$)
\begin{eqnarray}
  \tmmathbf{H}_l & = & A \nabla^2 \tmmathbf{m}_l + Km^y_l \hat{\tmmathbf{y}} +
  J\tmmathbf{m}_{\bar{l}},
\end{eqnarray}
such that
\begin{eqnarray}
  \delta E & = & - \sum_l \int \tmmathbf{H}_l \cdot \delta \tmmathbf{m}_l
  \mathd \tmmathbf{r}.
\end{eqnarray}
Since we require $\tmmathbf{m}_l \cdot \delta \tmmathbf{m}_l = 0$, the stable
configuration (denoted as $\bar{\tmmathbf{m}}_l$ and $\bar{\tmmathbf{H}}_l$) should satisfy $\bar{\tmmathbf{m}}_l \times \bar{\tmmathbf{H}}_l = \tmmathbf{0}$.

The Landau-Lifshitz equation governs the magnetic dynamics:
\begin{eqnarray}
  \gamma^{- 1} \frac{\mathd \tmmathbf{m}_l}{\mathd t} & = & -\tmmathbf{m}_l
  \times \tmmathbf{H}_l,
\end{eqnarray}
where $\gamma$ is the gyromagnetic ratio.

Since $\bar{\tmmathbf{m}}_l$ is gradually changing in the real space, we introduce spatial varying local coordinates, as explained in the main text:
\begin{eqnarray}
  \bar{\tmmathbf{m}}_l & = & (\sin \theta_l \cos \phi_l, \sin \theta_l \sin
  \phi_l, \cos \theta_l),\nonumber\\
  \delta \tmmathbf{m}_l & = & \delta m^{\theta}_l
  \hat{\tmmathbf{e}}^{\theta}_l + \delta m^{\phi}_l
  \hat{\tmmathbf{e}}^{\phi}_l,\nonumber\\
  \delta \tmmathbf{H}_l & = & \delta H^r_l \hat{\tmmathbf{e}}^r_l + \delta
  H^{\theta}_l \hat{\tmmathbf{e}}^{\theta}_l + \delta H^{\phi}_l
  \hat{\tmmathbf{e}}^{\phi}_l .
\end{eqnarray}
For a N\'eel type domain wall, we have $\theta_l (x) = \mathpi/2$.

With the following mathematical identities:
\begin{eqnarray*}
  \partial_x \hat{\tmmathbf{e}}_l^r & = & (\partial_x \phi_l)
  \hat{\tmmathbf{e}}^{\phi}_l,\\
  \partial \hat{\tmmathbf{e}}^{\phi}_l & = & - (\partial_x \phi_l)
  \hat{\tmmathbf{e}}_l^r,\\
  \partial^2_x \hat{\tmmathbf{e}}_l^r & = & - (\partial_x \phi_l)^2
  \hat{\tmmathbf{e}}_l^r + (\partial^2_x \phi_l)
  \hat{\tmmathbf{e}}^{\phi}_l,\\
  \partial^2_x \hat{\tmmathbf{e}}^{\phi}_l & = & - (\partial^2_x \phi_l)
  \hat{\tmmathbf{e}}_l^r - (\partial_x \phi_l)^2
  \hat{\tmmathbf{e}}^{\phi}_l,\\
  \hat{\tmmathbf{y}} & = & \sin (\phi_l) \hat{\tmmathbf{e}}_l^r + \cos
  (\phi_l) \hat{\tmmathbf{e}}^{\phi}_l,
\end{eqnarray*}
we derive the expression for $\bar{\tmmathbf{H}}_l$ as
\begin{widetext}
\begin{eqnarray}
  \bar{\tmmathbf{H}}_l = [- A (\partial_x \phi_l)^2 + K \sin^2 (\phi_l) + J \cos
  (\phi_{\bar{l}} - \phi_l)] \hat{\tmmathbf{e}}_l^r + [A (\partial^2_x
  \phi_l) + K \sin (\phi_l) \cos (\phi_l) + J \sin (\phi_{\bar{l}}
  - \phi_l)] \hat{\tmmathbf{e}}^{\phi}_l .
\end{eqnarray}
\end{widetext}
Therefore, we conclude the stable configuration for $\phi_l$ should satisfy
\begin{eqnarray}
  A (\partial^2_x \phi_l) + K \sin (\phi_l) \cos (\phi_l) + J \sin
  (\phi_{\bar{l}} - \phi_l) & = & 0.
\end{eqnarray}

At this point, we take advantage of the fact that the ground state satisfies $C_{2x}$ symmetry, such that $\phi_1=-\phi_2$. With $\delta \tm^{\theta(\phi)}_\pm = (\delta m^{\theta(\phi)}_1 \pm \delta m^{\theta(\phi)}_2)/2$, we can express the dynamical equation for $\delta \bm{\tm}$ as ($\phi = \phi_1$)

\begin{widetext}
\begin{eqnarray}
  \gamma^{- 1} \frac{\mathd \delta \tm_+^{\theta}}{\mathd t} & = & A
  \nabla^2 \delta \tm_+^{\phi} + K \delta \tm^{\phi}_+ \cos (2
  \phi),\nonumber\\
  \gamma^{- 1} \frac{\mathd \delta \tm_+^{\phi}}{\mathd t} & = & - A
  \nabla^2 \delta \tm_+^{\theta} - A (\partial_x \phi)^2 \delta
  \tm^{\theta}_+ + K \sin^2 (\phi) \delta \tm^{\theta}_+ + J \cos (2 \phi)
  \delta \tm^{\theta}_+ - J \delta \tm^{\theta}_+,\nonumber\\
  \gamma^{- 1} \frac{\mathd \delta \tm_-^{\theta}}{\mathd t} & = & A
  \nabla^2 \delta \tm_-^{\phi} + K \delta \tm^{\phi}_- \cos (2 \phi)
  - 2 J \delta \tm^{\phi}_- \cos (2 \phi),\nonumber\\
  \gamma^{- 1} \frac{\mathd \delta \tm_-^{\phi}}{\mathd t} & = & - A
  \nabla^2 \delta \tm_-^{\theta} - A (\partial_x \phi)^2 \delta
  \tm^{\theta}_- + K \sin^2 (\phi) \delta \tm^{\theta}_- + J \cos (2 \phi)
  \delta \tm^{\theta}_- + J \delta \tm^{\theta}_-. \label{MMmagnonall}
\end{eqnarray}
\end{widetext}

{As explained in the main text, for $\delta \bm{\tm}_+$, the potential is always a trapping one and gives rise to confined magnon channels. The trapping potential $U^\theta(x)$ and $U^\phi(x)$ (see main text for definition) is plotted in Fig.~\ref{fig:MMmagnon_supp}(a).} For $\delta \bm{\tm}_-$, if $\phi=\uppi/2$ across the domain wall, the potential is still a trapping one. For nontrivial $\phi(x)$, whether 1D magnon exists is not obvious from Eq. (\ref{MMmagnonall}). For bilayer CrI$_3$, we numerically find the 1D magnon channel for $\delta \bm{\tm}_-$, which is plotted in Fig.~\ref{fig:MMmagnon_supp}(b).

\section{Magnon dispersion of RR and MM domain walls}

\begin{figure}
\centering
\includegraphics[width=1.0\columnwidth]{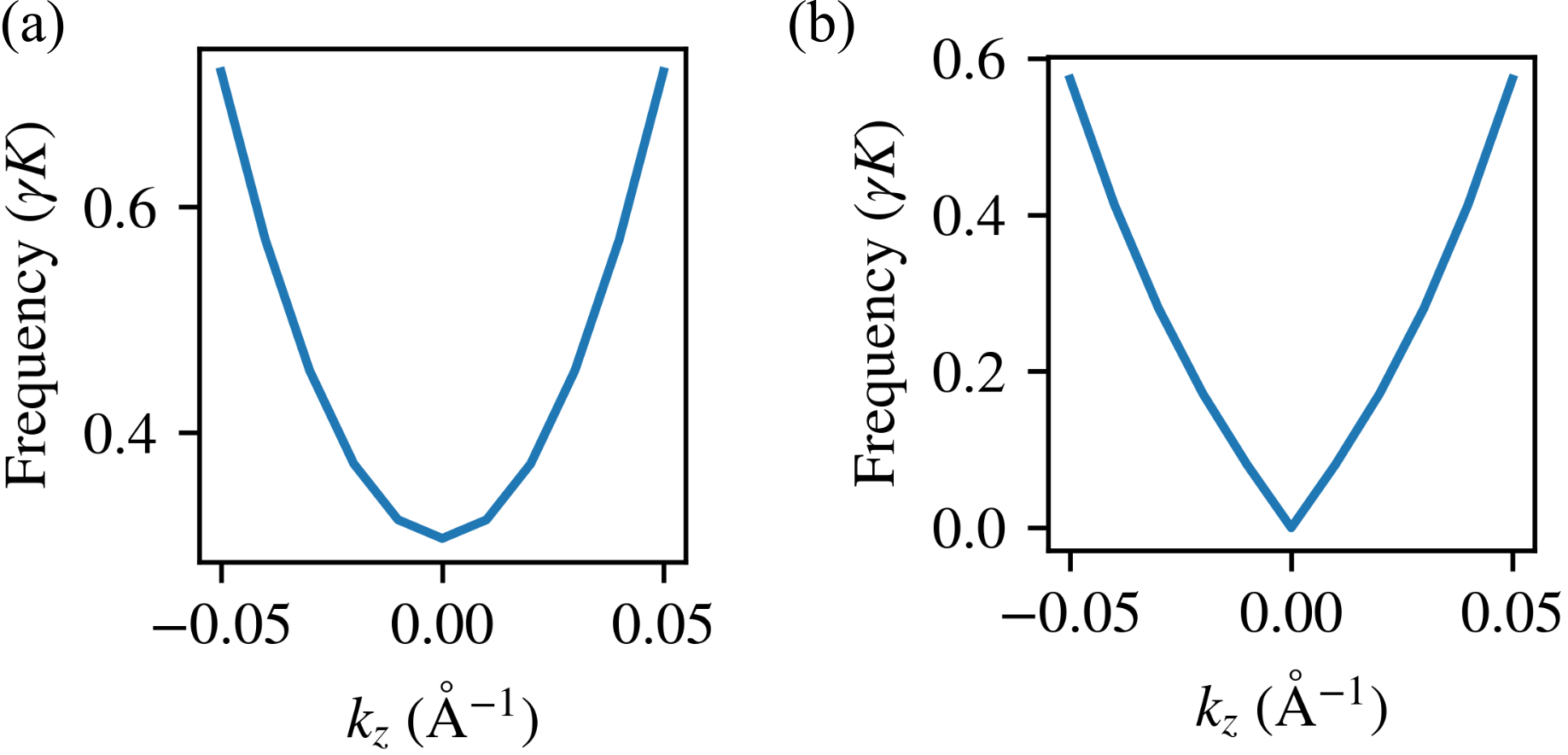}
\caption{{The dispersion of the 1D magnon modes of the lowest frequency for RR (a) and MM (b) domain walls. In both panels, $\gamma K\sim 0.3$ meV.}\label{fig:magnon_dispersion}}
\end{figure}

{Since our continuum model is translational invariant in the $z$ direction, the variation of magnetization unit vector $\delta \bm{m}$ propagates as plane waves in the $z$ direction, i.e., $\delta \bm{m} \propto \exp{(\mathi k_z z)}$. The frequency (alternatively energy) of the 1D magnons as a function of $k_z$ is plotted in Fig.~\ref{fig:magnon_dispersion}.

For the RR domain wall, it is clear from Eq.~(4) in the main text that the dispersion is quadratic. For the MM domain wall, Fig.~\ref{fig:magnon_dispersion}(b) shows the magnon of the lowest frequency (belonging to $\delta \tilde{\bm m}_+$) is linear for small $k_z$.}

\section{Excitation of the 1D magnons confined in the stacking domain walls}

\begin{figure}
\centering
\includegraphics[width=1.0\columnwidth]{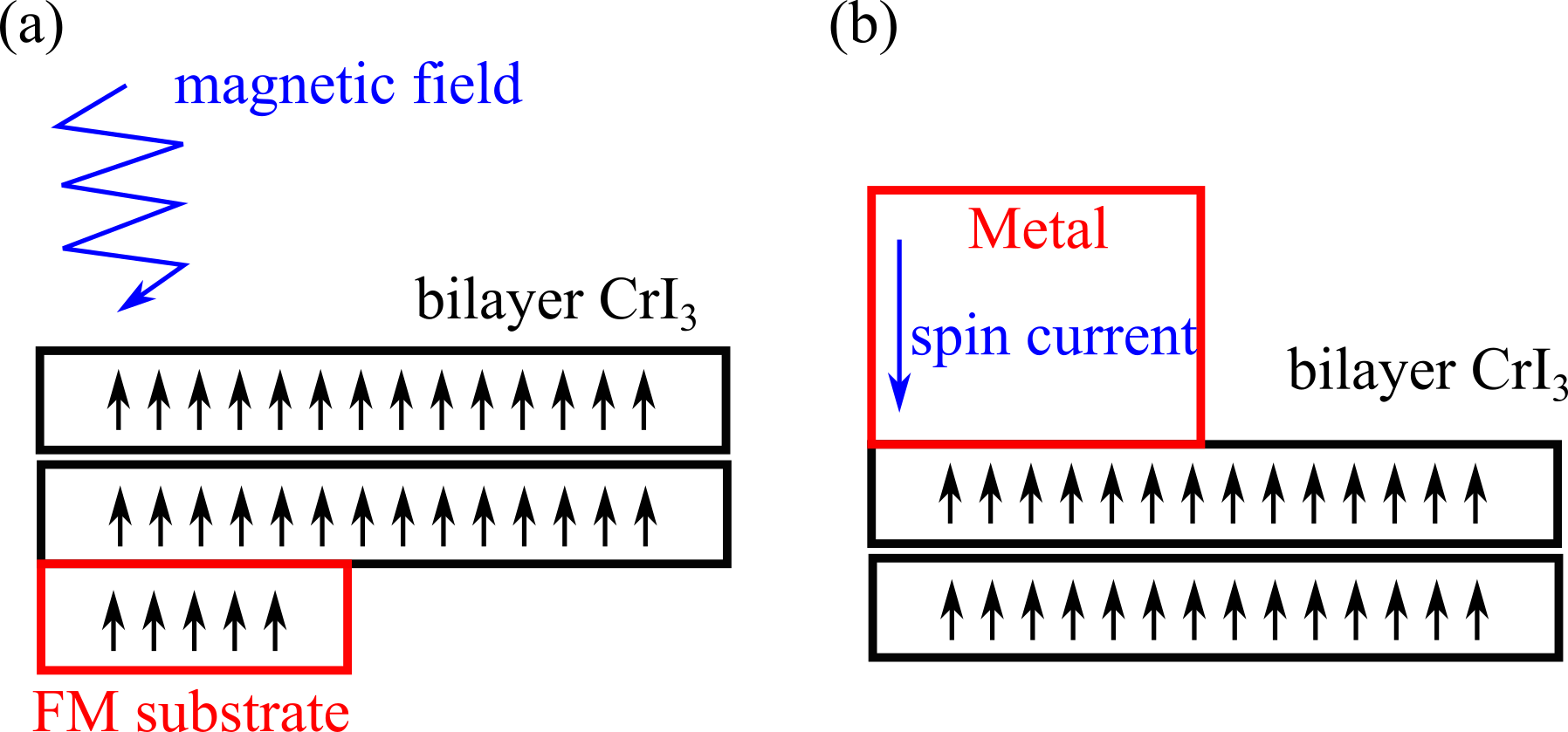}
\caption{{(a) Excitation of 1D magnon in the RR domain wall with an FM substrate. Note that the Rhombohedral stacking favors ferromagnetic interlayer configuration. (b) Excitation of 1D magnon in the RR domain wall with spin current. }\label{fig:excitation}}
\end{figure}

{For most spintronics applications, magnons need to be excited and detected. In the main text, we have mentioned that some of the magnons (e.g., the 1D $\delta \bm{m}_-$ magnon in RR domain wall) do not carry net magnetic moments. These magnons cannot be directly excited by a microwave magnetic field. In this section, we discuss several possibilities to excite these magnons:
\begin{itemize}
    \item Thermal excitation. As we discussed in the main text, the $\delta\bm m_+$ mode in the RR domain wall is not affected by the stacking domain wall, and thus it has the same energy as the bulk magnon ($\sim$0.3 meV). On the other hand, the 1D magnon (the $\delta\bm m_-$ mode) has an energy $\sim$0.09 meV. Thus it is the first populated magnon mode at finite temperature.
    
    \item Ferromagnetic proximity effect [cf. Fig.~\ref{fig:excitation}(a)]. One can try to place \emph{part of} the bilayer CrI$_3$ on a ferromagnetic substrate such that the magnetization of the bottom layer is more difficult to rotate. A microwave magnetic field can be used to excite a magnetic state that has nonzero overlap with the $\delta \bm m_-$ magnon mode. The magnetic state can propagate to places where the ferromagnetic substrate is absent, which is described by our theory.
    
    \item Spin convertance at magnetic interfaces~\cite{zhang_spin_2012} [cf. Fig.~\ref{fig:excitation}(b)]. One can put a suitable metal on top of the bilayer CrI$_3$. With a spin current denoted as the blue arrow in Fig. \ref{fig:excitation}(b), magnons can be excited. Since the excitation is through exchange coupling between the electrons of the metal and the magnetic moments in the top layer, the excited magnetic state will have nonzero overlap with the $\delta \bm m_-$ magnon mode.
\end{itemize}

Generally, besides the 1D magnons studied in this work, the above approaches will also excite bulk magnons. However, since bulk magnons propagate in all directions, they decay much more quickly than the 1D magnons. Therefore, the 1D magnons will be spontaneously separated from the bulk magnons in the propagation.

Finally, we note that some 1D magnons (e.g., the $\delta \tilde{\bm{m}}_+$ magnon mode in the MM domain wall) carry net magnetic moments and can be excited with a microwave magnetic field.
}

\bibliography{main}